\documentclass[12pt,  reqno]{amsart}
\usepackage{graphicx,amsmath,amssymb,mathtools,amsthm}
\usepackage{color}
\usepackage{natbib}
\usepackage{hyperref}
\hypersetup{colorlinks=true,
	citecolor = blue,
	filecolor=cyan}
\usepackage{cleveref}
\usepackage{float}
\usepackage{bbm}
\usepackage{longtable}
\usepackage{caption}
\usepackage{subcaption}
\usepackage{enumitem}
\usepackage{appendix}
\usepackage{booktabs}
\usepackage{pdflscape}

\usepackage[left=1in,right=1in, top=1in, bottom=1in]{geometry}

\long\def\comment#1{}

\newtheorem{proposition}{Proposition}

\theoremstyle{definition}

\newtheorem{remark}{Remark}

{\title[]{The learning effects of subsidies to bundled goods: a semiparametric approach $^*$}}

\author[]{Luis Antonio F. Alvarez $^\dag$}

\address{ Department of Economics, University of S\~{ao} Paulo, 05508-010 S\~{ao} Paulo, Brazil.}
\email{\href{mailto:luisfantozzialvarez@gmail.com}{luisfantozzialvarez@gmail.com}}

\author[]{Ciro Biderman}

\address{ S\~{ao} Paulo School of Business Administration, 01313-902 S\~{ao} Paulo, Brazil.}
\email{\href{mailto:ciro.biderman@fgv.br}{ciro.biderman@fgv.br}}
\date{May, 2024.}

\thanks{* This paper subsumes and supersedes our earlier unpublished paper ``Semiparametric analysis of randomised experiments using L-moments''. The authors are grateful to Ricardo Masini, Victor Orestes, Cristine Pinto, Pedro Sant'Anna and participants at the 44th Brazilian Econometric Society Meeting and FEA-USP seminar series for their useful comments and suggestions. This paper relies on experimental data from randomized control trial AEARCTR-0003518. Biderman gratefully acknowledges Nathalie Picarelli, Miguel Jacob, Bianca Bianchi, Ana Guerrini, Leonardo Bueno and Rodrigo Ferreira for their work in implementing the trial and in collecting post-intervention data. Alvarez gratefully acknowledges financial support from Fapesp grant 2022/16691-1. Biderman gratefully acknowledges support from Fapesp grant 2021/11959-3.}

\thanks{$\dag$ Corresponding author. Department of Economics, University of S\~{ao} Paulo, 05508-010 S\~{ao} Paulo, Brazil. Phone number: +5513996265171. E-mail address: \href{mailto:luisfantozzialvarez@gmail.com}{luisfantozzialvarez@gmail.com}}

\makeatletter
\newcommand{\addresseshere}{%
	\enddoc@text\let\enddoc@text\relax
}
\makeatother

\begin{document}

\thispagestyle{plain}
\begin{abstract}
   Can temporary subsidies to bundles induce long-run changes in demand due to learning about the quality of one of the constituent goods? This paper provides theoretical support and empirical evidence on this mechanism. Theoretically, we introduce a model where an agent learns about the quality of an innovation through repeated consumption. We then assess the predictions of our theory in a randomised experiment in a ridesharing platform. The experiment subsidised car trips integrating with a train or metro station, which we interpret as a bundle. Given the heavy-tailed nature of our data, we propose a semiparametric specification for treatment effects that enables the construction of more efficient estimators. We then introduce an efficient estimator for our specification by relying on L-moments. Our results indicate that a ten-weekday 50\% discount on integrated trips leads to a large contemporaneous increase in the demand for integration, and, consistent with our model, persistent changes in the mean and dispersion of nonintegrated app rides. These effects last for over four months. A calibration of our theoretical model suggests that around 40\% of the contemporaneous increase in integrated rides may be attributable to increased incentives to learning. Our results have nontrivial policy implications for the design of public transit systems. 
   
   \vspace{0.5em}
   
   \vspace{0.5em}
\end{abstract}

\maketitle

\pagebreak
\section{Introduction}

In markets for innovations, it is reasonable to assume that consumers may be uncertain about product quality. In these settings, firms may offer a bundle, which combines an innovation with other goods of better known quality, in order to mitigate consumer uncertainty and induce \textit{learning} with respect to the novelty \citep{Reinders2010}. Typical examples include packaging trial versions of new software with updates of better known software \citep{Sheng2009}, and the bundling of financial innovations with consolidated services \citep{tufano2008using}.

In these innovation settings, bundles are often offered at a discount. Such \emph{subsidy} is expected to increase learning about the innovation. Indeed, inasmuch as substitution effects dominate, we may expect users to increase their demand of the bundled good in response to the subsidy, thereby acquiring additional information about the innovation. Moreover, a forward-looking agent may increase demand due to an additional \textit{experimentation} or \textit{learning motive}, i.e. in order to increase the informational bequest to be left for her future selves.

Do temporary subsidies to bundled goods induce long-term changes in consumption behaviour due to increased learning on the relative quality of one of the constituent goods? This paper provides theoretical guidance and empirical evidence on the role of this mechanism. Theoretically, we introduce a dynamic model where an agent learns about the quality of an innovation on an essential good through repeated consumption. The agent has a target level of consumption of an essential good, as well as preferences over consumption in other goods. The target can be fulfilled by relying on a known outside option, an innovation whose relative quality is uncertain, or a bundle with a share $\phi$ of the outside option and $1-\phi$ of the unknown good. Increasing consumption of either the bundle or the innovation increases signal precision regarding the underlying quality, albeit at possibly different rates. In this setting, we show that the effect of a one-period subsidy on the bundle can be decomposed into a direct price effect, and indirect effects due to a \textit{learning or experimentation motive}, as the agent leverages the discount to increase the informational bequest to be left to her future selves. Our results further show that subsidies may affect not only the mean, but also the dispersion of demand, and that, given the essentiality nature of the innovation, any extemporaneous effects of a one-period subsidy to the bundle are fully attributable to learning.

Empirically, we assess the predictions of our theory of learning in the context of a randomised experiment conducted in partnership with a large ridesharing service in S\~{a}o Paulo, Brazil. The experiment randomly provided two ten-day discount regimes to a subset of the platform's user base. One of the regimes provided two daily 20\% discounts, limited to 10 BRL per trip (around 4 USD at the time), on rides starting or ending at a train/metro station (\textit{integrated rides}). The other regime provided two daily 50\% discounts, also limited to 10 BRL per trip, on integrated rides. We also have access to a control arm, which was not informed about the experiment. For each individual in the experimental sample, we observe baseline survey answers on socioeconomic information and commuting habits, as well as the user history of demand for integrated and \textit{non-integrated} (total minus integrated) rides on a fortnightly basis, starting four months prior to the experiment and going over six months after the discount ends. We view this setting as an ideal environment to test the predictions of our theory, as we observe a one-off unanticipated discount to a bundled good (integrated rides), which may generate information about the underlying quality of an innovation (non-integrated rides)\footnote{Ridesharing apps were authorised to operate in São Paulo in mid 2016; and the experiment was conducted in late 2018.} on an essential good (transportation services). We are also able to assess long-term effects of discounts, given the time range spanning over six months after the experiment took place.

The distribution of individual demand for rides exhibits heavy tails, which leads to inferences based on difference-in-means estimators performing poorly \citep{Lewis2015}. Following \cite{Athey2023}, we propose a semiparametric model for potential outcomes that, while constraining treatment effect heterogeneity, enables the construction of more efficient estimators. Motivated by our theory, we consider a parametrisation that allows treatment effects to change both the location and the scale of the demand for rides. By stratifying our analysis, we also allow for heterogeneity according to user's information sets. We then introduce a semiparametric estimator for our proposed method by relying on L-moments, a robust alternative to standard moments that characterise any distribution with finite mean \citep{Hosking1990}. Our estimator, a semiparametric version of the generalized method of L-moments estimator of \cite{alvarez2023inference}, has several attractive properties: in our adopted parametrisation, it can be easily computed as a generalised least squares estimator in a modified dataset; it yields as an immediate byproduct a specification (overidentifying restrictions) test of the adopted model; and, in the setting of \cite{Athey2023}, it is asymptotically efficient without requiring any additional corrections, such as estimation of the model's efficient influence function \citep{Klaasen1987,Newey1990}.

When applied to our experimental data, the semiparametric methods shed light on several patterns. The 50\% discount leads to a large contemporaneous increase in integrated rides, amounting to 60\% of the control group average at the period. However, there do not appear to be subsequent effects in either the mean or dispersion of integrated rides. As for the demand for non-integrated rides, we observe a sizeable and persistent decrease in their mean and dispersion in the fortnights after the discount took place. These effects last for over four months after the end of the experiment. Reductions in average non-integrated rides amount to 15\% of the control group mean three months after the discount. We observe similar, albeit weaker, effects on the 20\% discount regime, though in this case the contemporaneous change in average integrated rides is not statistically significant. 

As we argue later on, observed patterns are mostly consistent with our learning model. Indeed, we may interpret our results through the following mechanism of our theory: upon increasing consumption as a response to the discount, agents increase learning about the relative (to the outside option) quality of ridesharing apps, updating their beliefs to a more pessimistic level. This reduces the demand for non-integrated rides. However, given the essentiality nature of transportation services, the consumer must shift consumption to other modes of transportation in order to meet her target. This substitution effect approximately compensates the impact of the pessimistic update on integrated rides -- recall the effect of quality on the bundle is dampened by the share $\phi$ --, and we do not observe changes in the demand for this type of trip after the discount is over. Note that, as a consequence of this mechanism, we would also expect an increase in the demand for the (unobserved) outside option in subsequent periods. Finally, to shed more light on the learning mechanism, we combine our experimental estimates with survey data to calibrate some key parameters of our model of learning. We find that, for the 50\% discount, around 40\% to 50\% of the contemporaneous increase in integrated rides may be attributed to the experimentation motive, i.e. to a desire to accumulate more information on transportation services for future use.

Our paper is connected to a larger body of research that seeks to provide a rationale for the existence of bundled goods. Typical explanations range from cost-saving and preference-complementarity arguments \citep{Pyndick}, to the use of bundles as a means of implementing price discrimination \citep{Adams1976}, as well as a mechanism used by incumbent firms to deter market entry \citep{carlton2002strategic,nalebuff2004bundling}. Firms may also offer bundles as a means of extracting surplus when product search is costly to consumers \citep{Harris2006}, and firms have access to a technology that is able to predict those goods most suitable to a consumer, e.g. recommender systems in streaming services \citep{hiller4415301digital}. In this paper, we provide theoretical as well as empirical evidence on yet another role played by bundles: they may be used to mitigate consumer uncertainty regarding an innovation and induce long-run changes in consumer behaviour via learning effects. Furthermore, by providing a formal theoretical foundation on the learning effects of subsidies to bundles and in-the-field experimental evidence on the role of this mechanism, we contribute to a body of research in marketing and product development that previously relied on laboratory and observational data to discuss bundling strategies for new product introduction \citep{SIMONIN1995219,Sheng2009,Reinders2010}.

The present work is also related to the literature on the markets and pricing policies of experimentation goods \citep{shapiro1983optimal,liebeskind1989markets,bergemann2006dynamic,Chen2022}. Experimentation goods are commodities whose underlying quality is only (partially) revealed after consumption. Our paper provides evidence that bundling, when coupled with a temporary subsidy, may be a convenient strategy to induce long-run changes in demand for this class of goods.\footnote{In a two-period setting, \cite{shapiro1983optimal} shows that, if consumers initial expectations regarding the quality of the experimentation good are pessimistic, the optimal strategy of a monopolist would be to provide an initial subsidy to increase adoption. In such situation, our results suggest that bundles may offer a complementary, and possibly cheaper, approach to induce these desired changes.} The nature of our experimental variation is such that our results follow without any assumptions on market equilibrium or firm behaviour. As a consequence, we view our paper as documenting a mechanism in consumer behaviour that may be incorporated in future research on the optimal design of pricing policies in these settings.

From the econometric standpoint, our paper contributes to the literature on semiparametric estimation \citep{Newey1990,Bickel1993,Newey1995,Bolthausen2002,kosorok2008introduction}. We introduce an alternative estimator for \citeauthor{Athey2023}'s model that is asymptotically efficient without requiring any further adjustments, e.g. the correction by a cross-fitted estimate of the model efficient influence function employed by \cite{Athey2023}. Our estimator is computationally convenient; and its objective function immediately provides us with a J-test of over-identifying restrictions. More generally, we view our paper as introducing the generalized method of L-moments approach of \cite{alvarez2023inference} -- which, in a parametric setting, improves upon maximum likelihood estimation in finite samples of popular distributions whilst remaining asymptotically equivalent to it --, to a semiparametric environment.\footnote{\cite{alvarez2023inference} consider an extension of their main approach that can be used to fit a parametric model for the error term of a semiparametric model that is estimated in a first step. Their proposed construction and the range of applications covered by their method are quite distinct from ours, though. In particular, their method requires estimating an influence-function correction to account for first-step estimation, whereas our method does not rely on knowledge or estimation of any influence function.} Given the attractive statistical properties of L-moments (see \Cref{sec_estimation} and  Online \Cref{app_metrics} for a discussion; also \citeauthor{Alvarez2023_mixture}, \citeyear{Alvarez2023_mixture}), we view this paper as a further step in developing generalised method of L-moment estimation in a semiparametric setting.\footnote{In a recent paper, \cite{Alvarez2023_mixture} show that a sieve-like extension of the estimator in \cite{alvarez2023inference} provides a computationally convenient estimator with valid inferential guarantees in \textit{nonparametric quantile mixture models}.}

Finally, our paper is also connected to the literature on the Economics of Urban Transportation. Methodologically, we introduce a dynamic model of demand for transportation services, which contrasts with the discrete choice models typically adopted in the literature \citep{small2007economics}. Our modelling approach may be more broadly useful to generate predictions in settings where learning is an important component, and individual choices are observed as aggregates over repeated time windows. Our paper also contributes to the literature on behavioural interventions in transportation markets \citep{METCALFE2012503,Kristal2019,GRAVERT20211}. In both developing and developed countries, a long-run declining trend in the use of public trasnportation services has been documented \citep{mallett2018trends,Rabay2021}. This decline has been further accentuated by the Covid-19 emergency \citep{mallett2022public,loh2023ensuring}, with a full recovery to pre-pandemic patterns very much unlikely absent further intervention \citep{Dai2021,Tsavadari}. Given the high fixed costs of operating a public transportation system, these changes tend to threaten the long-run financial viability of these services, which often translates into increased government subsidies flowing into the operation \citep{welle2020safer,aguilar2021after,Tsavadari}. Insofar as the outside option in our analysis may be interpreted as public transportation,\footnote{As we argue in more detail in \Cref{background}, we restrict our empirical analysis to a subsample of users for which the outside option is more likely to be interpreted as the train/metro system. Moreover, we note that, in our subsample of interest, public transportation usage is quite prevalent, with 68.8\% of participants reporting the train, metro or public bus system as one of their main modes of transportation.} our results would indicate that temporary subsidies to modal integration may induce persistently higher public transportation takeup by improving beliefs regarding the relative quality of the service.\footnote{Public transportation trips would increase because the average number of integrated rides does not decrease and, when interpreting the results of our experiment through the lenses of our theory, we expect an increase in the outside option after the discount is over.} This result suggests that these types of policies could be employed to partially offset long-run trends in commuting behaviour. 

The remainder of this paper is organised as follows. Section \ref{model} introduces our model. Section \ref{background} discusses the experimental design. Section \ref{identification} presents our target estimands and the proposed semiparametric model. Section \ref{sec_estimation} discusses estimation. Section \ref{results} presents the results of our empirical analysis. Section \ref{conclusion} concludes. The \hyperref[supplementary]{Online Appendix} contains the proofs of our main results, as well as additional details on the semiparametric estimator.

\section{Model}
\label{model}
In this section, we introduce a model which aims to capture the main mechanisms associated with discounts to bundled goods in a setting where the quality of one of the elements in the bundle is unknown. We focus on an environment where the goods of unknown quality are \emph{essential}, in the sense that the consumer has a target level of consumption required in order to be able to consume other goods. Our leading example are transportation services, which are the focus of our empirical application. It seems reasonable to assume that preferences for commuting services arise due to having to work or study outside home, which generates a target level of transportation needs that must be met.\footnote{Indeed, this approach is in keeping with both textbook discussions on the demand for transportation services \citep{small2007economics}, as well as more recent treatments \citep[e.g.][]{Kreindler2023}.}

Consider a consumer which, at every period $t \in \mathbb{N}$, must decide between three modes of transportation: a known outside option ($o_t$), a new mode of transportation  ($n_t$), and an integrated trip ($b_t$), which is a bundle consisting of the two types of transportation services. The amount consumed of each type of ride is then combined to produce total transportation services ($\tau_t$), according to the technology:

$$\tau_t = o_t + \gamma A^\phi b_t + A n_t \, , $$
where $A\geq 0$ is the relative quality of the new mode of transportation. The parameter $\phi \in (0,1)$ captures the fact that the new mode of transportation enters the composition of an integrated trip.\footnote{Indeed, the term $\gamma A^\phi b_t$ may be seen as a reduced form for the problem of choosing the composition of an integrated trip between the outside option and the innovation with Cobb-Douglas technology.}

Relative quality $A$ is unknown, but, at period $t$, the consumer has information $\mathcal{H}_t$ over it, where $\mathcal{H}_t$ is a sub-$\sigma$-algebra of $\mathcal{S}$, with $(S, \mathcal{S},\mathbb{P})$ being a probability space prescribing the uncertainty regarding $A$.\footnote{In other words, $A$ is a random variable defined on $(S, \mathcal{S},\mathbb{P})$.} We remain agnostic about the nature of uncertainty encoded in $\mathbb{P}$ -- it may reflect objective or subjective factors -- though we assume the consumer is an expected utility maximiser. Specifically, we assume that the consumer has preferences over both transporation services ($\tau_t$) and other forms of consumption ($c_t$), and that, given information at $t$, his expected utility at period $t$ is given by:

\begin{equation*}
    u_t(c_t,\tau_t|\mathcal{H}_t) = \mathbb{E}[U_t(c_t,\tau_t)|\mathcal{H}_t] = c_t - \frac{\lambda}{2}  \mathbb{E}[(\tau_t - \tau^*_t)^2|\mathcal{H}_t] \, ,
\end{equation*}
where  $\lambda > 0$ and $\tau_t^*\geq 0$. In this setting, transportation enters utility as an \emph{essential} good: the consumer has a target level of transportation ($\tau^*_t$), which may be thought as the amount required for receiving her income (or education), and any amount above or below that is undesirable to her.\footnote{The assumption of linearity in consumption may be seen to reflect that, in the effective range where the choice between transportation modes is made, the marginal utility of consumption of other goods is constant. See \cite{Argente2022} for a similar assumption of quasilinearity in outside consumption in a model for the demand for Uber rides under alternative payment methods.}

We are now ready to describe the dynamics of the choice problem. After accruing transportation services, the consumer receives a signal of the underlying quality $A$. We assume that this signal, $Z_t$, is given by $$Z_t = \psi(A) + \frac{1}{\bar{h}(b_t,n_t)}  V_t, $$ where  $\psi$ is a bimeasurable one-to-one map; $V_t|\mathcal{H}_t,A$ follows a stable distibution function $F$ with zero mean and unit variance:\footnote{In our setting, the distribution function $F$ is stable if, for every $c,c' \in \mathbb{R}$ and independent copies $Z, Z'$ distributed as $F$, there exists $c''$ such that $cZ+c'Z' \overset{d}{=} c'' Z$.  } and $\bar{h}$ is a smooth increasing nonnegative function. This formulation captures the idea that, the more the user utilises the innovation, the more precise is the signal about its underlying quality. After observing the signal $Z_t$, information is then updated as:

$$\mathcal{H}_{t+1} = \sigma(\mathcal{H}_t, Z_{t}) \, .$$

Finally, given a discount factor $\beta \in (0,1)$, a consumer's value at $t$ is given by:

\begin{equation}
\label{eq_model}
\begin{aligned}
        V_t(a_t| \mathcal{H}_t) = & \max_{o,b,n,c\geq 0, a' \geq \underline{a}_{t+1}} u_t(c,\tau| \mathcal{H}_{t+1}) + \beta \mathbb{E}[V_{t+1}(a'|\mathcal{H}_{t+1}(\bar{h}(b,n))|\mathcal{H}_t]   \, ,
\end{aligned}
\end{equation}
subject to the technology and budget constraints:

\begin{equation}
    \begin{aligned}
    \tau = o  + \gamma A^\phi b + n \, , \\
        p_{o,t} o + p_{b,t} b + p_{n,t} n + c + a' \leq (1+i_t)a_t + w_t \, ,
    \end{aligned} 
\end{equation}
where $a_t$ is the agent's wealth, and $w_t$ is her (non-interest) income, and $i_t$ is the interest rate. The notation $\mathcal{H}_{t+1}(\bar{h}(b,n))$ represents the $\sigma$-algebra $\sigma(\mathcal{H}_{t}, \psi(A) + \frac{1}{\bar{h}(b,n)} V_t)$

The following proposition collects some properties of the learning model. In what follows, denote by $(a_t,c_t,b_t,n_t,o_t)_{t \in \mathbb{N}}$ the optimal choices of wealth and consumption: 

\begin{proposition}
\label{prop_learning}
   Consider the learning model previously outlined. Assume that $\mathbb{E}[A^2] < \infty$. Fix an initial information set $\mathcal{F}_1 \subseteq \mathcal{S}$, initial wealth $a_1 \geq \underline{a_1}$, and consider bounded sequences $(w_t)_{t \in \mathbb{N}}$ and $(\tau^*_t)_{t \in \mathbb{N}}$. Suppose that $ \beta (1+i_t) \leq 1$ for every $t \in \mathbb{N}$; and that $\underline{a}_t$ is greater or equal than the natural debt limit at $t$. We then have that:

    \begin{enumerate}
        \item For every $t \in \mathbb{N}$, the expected future value of having a more precise signal is always nonnegative, i.e. for any $h'', h' \in \operatorname{Im}(\bar{h})$, $h'' > h'$:

        \begin{equation*}
            \mathbb{E}[V_{t+1}(a_{t+1},\mathcal{H}_{t+1}(h''))|\mathcal{H}_t] \geq      \mathbb{E}[V_{t+1}(a_{t+1},\mathcal{H}_{t+1}(h'))|\mathcal{H}_t]  \, .
        \end{equation*}
        \item If there are sufficient incentives to learning, inasmuch that, at an optimum:

        $$\limsup_{n \to \infty}\frac{n}{\sum_{j=1}^n \bar{h}(b_j,n_j)} < \infty \, \text{a.s.} ,$$
        then, asymptotically, the consumer achieves perfect knowledge of the underlying quality, in the sense that, for any measurable function $g$ such that $\mathbb{E}[|g(A)|]< \infty$:
        \begin{equation*}
           \lim_{n \to \infty} \mathbb{E}[g(A)|\mathcal{H}_n] \overset{a.s.}{\to} g(A) 
        \end{equation*}
        and $\mathbb{E}[|\mathbb{E}[g(A)|\mathcal{H}_n] - g(A)|] \to 0 $.
        
        \item  In an optimum with positive levels of consumption of each type of good at each period, the effect of an unanticipated, small\footnote{Small such that the new equilibrium still features positive consumption of all goods.} one-period  change in the price of integrated rides on the demand for different modes of transportation is:
        \begin{enumerate}
            \item contemporaneously, it can be decomposed as:
            \begin{equation*}
                \begin{bmatrix}
                   \frac{\Delta o_t}{\Delta p_{b,t}} \\                         \frac{\Delta b_t}{\Delta p_{b,t}} \\
                           \frac{\Delta n_t}{\Delta p_{b,t}}
                \end{bmatrix} = \frac{\boldsymbol{\phi_{1,t}}}{\lambda} + \frac{\boldsymbol{\phi_{2,t}}}{\lambda} \frac{\Delta \text{ marginal value of learning from } b_t}{\Delta p_{b,t}} + \frac{\boldsymbol{\phi_{3,t}}}{\lambda} \frac{\Delta \text{marginal value of learning from } n_t}{\Delta p_{b,t}} \, ,
            \end{equation*}
            where $\boldsymbol{\phi_{1,t}}$, $\boldsymbol{\phi_{2,t}}$ and $\boldsymbol{\phi_{3,t}}$ are elements of $\mathbb{R}^3$ that depend on $\gamma$ and the conditional moments $\mathbb{E}[A^s|\mathcal{H}_t]$, $s\in \{\phi, 1, 2 \phi, \phi+1, 2\} $.
            \item in future periods, it is \textbf{fully} mediated by learning, in the sense that, if the information set $\mathcal{H}_{t+1}$ were to remain unchanged after a discount at period $t$, then there would be no changes in demand for transportation in periods $s > t$ relative to the no discount case.
        \end{enumerate}
    
    \end{enumerate}

    \begin{proof}
        See Online \Cref{proof_prop}.
    \end{proof}
\end{proposition}

The first part of the proposition shows that information is always valued in the model. In a finite horizon setting, a proof of this fact (for every period up to the terminal one) follows from extensions of Blackwell's comparison of experiments theorem \citep{Blackwell1953,DeOliveira2018} to general state spaces \citep{khan2020missing}. In our infinite horizon setting, we alternatively rely on the stability assumption on the signal coupled with boundedness of income and the target to offer a direct proof.

The second part of the proposition relies on known results on martingales \citep{Durrett2019} to show that if, in an equilibrium, there are enough incentives to learning, inasmuch that the signal has always a sufficient degree of informativeness, then, asymptotically, the consumer is able to perfectly recover the underlying quality. An implication of this fact is that, in a world featuring consumers heterogeneous with respect to their beliefs and preferences, but a common underlying quality $A$ entering utilities, then, asymptotically, heterogeneity in consumption patterns due to different information vanishes.

The third part of the proposition is concerned with the effects of a one-period small change in the price of integrated rides on the demand for transportation. Item 3.(a) shows that, contemporaneously, this effect can be decomposed into a direct price effect, and indirect effects mediated by changes in the marginal value of learning. The latter may be interpreted as a change in the \emph{learning or experimentation motive for demand}, as a one-period change in the price of integrated rides also alters incentives to accumulate information for the agent's future selves. The magnitude of each component is determined by the consumer's information regarding $A$ at the beginning of the period, as well as the technology parameters for bundled goods, $\gamma$ and $\phi$. Item 3.(b) shows that, in our model, extemporaneous effects on the demand for transportation are fully mediated by learning. Put another away, if a discount were not to induce any additional learning, then there would be no changes in the demand for transportation other than in the period where the discount takes place. Such phenomenon is driven by the essentiality nature of public transportation in the model.\footnote{Indeed, if there was complementarity between consumption and transportation in the model (e.g., the utility of consumption is given by $c_t^\alpha \tau_t^{1-\alpha}$ for some $\alpha \in (0,1)$), then intertemporal substitution effects in $c_t$ may induce changes in the demand for public transportation in other periods \emph{even if} there was no change in the information acquired.}

We summarise below the two main predictions from the learning model, which will guide our empirical analysis:
\begin{enumerate}
    \item  In a world with heterogeneous agents, if every consumer experiences the same relative quality $A$, then, under sufficient incentives to learning, asymptotically heterogeneity in consumption patterns due to different signals or prior beliefs vanishes. As a consequence, increasing incentives to learning in this setting is expected to produce changes in the dispersion of the demand for transportation services.
    \item The extemporaneus impact of a temporary reduction in the relative price of transport services is fully mediated by learning.
\end{enumerate}

\section{Background and data}
\label{background}
We assess the predictions of our theory of learning in the context of transportation markets in S\~{a}o Paulo, Brazil.  S\~{a}o Paulo is Brazil's largest municipality, with around 11.5 million inhabitants reported in the 2022 national Census \citep{census2022}. Its transportation infrastructure includes a metro system extending short over 100km \citep{metro2022}, nearly 200km of commuter rail lines connecting the municipality to its larger metropolitan area \citep{cptm2022}, 1,300 bus routes \citep{sptrans2023} and 700 km of bus lanes \citep{sptrans2022}, and 722 km of bicycle lanes \citep{cet2022}. Since 2016, ride-hailing services are authorised to operate by the local government upon complying with regulations \citep{g12016,biderman2020regulating}.

We rely on data from a randomised experiment which one of the authors conducted in the municipality of S\~{a}o Paulo \citep{Biderman2018} in partnership with a company that started operating a ridesharing service locally in 2016.\footnote{Before 2016, the company operated a local cab calling service. Cab rates are fixed by the local government, whereas ridesharing fees are dynamically set by the platform's algorithm.} In 2018, this service implemented an experiment among its registered users with an aim to understand the potential for its app rides being used to complete the first- or last-mile of a train/metro trip. The experimental design proceeded in two steps. In the first step, the company sent a message to a random sample of 60,000 of its users asking them to answer a survey in exchange for a 15 BRL\footnote{Around 4 USD at the 2018 exchange rate.} discount coupon upon completion. The initial survey collected socioeconomic data, as well as baseline information on commuting habits.\footnote{The variables collected in the baseline survey are presented in Table \ref{table_data_raw} in the Online Appendix.} At this time, individuals were only aware of the 15 BRL coupon, with no further information on a future experiment being disclosed. 

In a second step, a randomised experiment was conducted among the survey respondents. The experiment consisted in providing discounts to app rides starting or ending at a subway/train station during two weeks between the end of November and the beginning of December 2018. Respondents were randomly divided into three arms: (i) a control group (which was not informed about the experiment); (ii) a group eligible to two 20\% discounts per day on rides starting/ending at a train/metro station,\footnote{The e-hailing company defined a virtual fence around the station. If the trip started or ended inside this fence, it was eligible to the discount.} limited to 10 BRL per ride; and (iii) a group eligible to two 50\% discounts per day on rides starting/ending at a train/metro station, also limited to 10 BRL per ride. Announcement of the discounts proceeded as follows. On November 26th, 2018 (a Monday), users selected into treatment received SMS and push notifications informing them of the availability of two daily coupons for rides integrating with the train/metro system until November 30th (same week’s Friday). They also received daily reminders for the rest of the business week. On the next Monday (December 3rd), all users in these groups received similar notifications informing them that the discounts were also available for that whole week (until December 7th, Friday), again with subsequent daily reminders. As a consequence, users assigned treatments status were eligible for up to 200 BRL in discounts during these two weeks.

For each respondent, we observe their baseline survey answers, as well as the number of \emph{integrated} (starting/ending at train/metro station) and \emph{non-integrated} (total minus integrated) app rides on a fortnightly basis, from early 2018 until mid 2019. We view this setup as an ideal environment to assess the predictions of our theory. Indeed, it appears reasonable to model transportation as an essential good. Moreover, from the lenses of the model in \Cref{model}, we observe a one-off unanticipated discount to a bundled good (integrated rides), which upon utilization may provide information about a new mode of transportation (non-integrated rides). Finally, our time window, which spans over six months after the discount takes place, allows us to evaluate the long-run effects of this short-term discount.

To further improve the interpretation of our results, we restrict our analysis to respondents who reported either living, working or studying close to a train/metro station. We do so for two main reasons. First, it better allows us to interpret learning about $A$ as ``learning about the relative quality of the innovation, \emph{vis-à-vis the train-metro option}''. Indeed, users who do not report being usually close to a train-metro station may combine the subsidy with other modes of transportation (e.g. renting a bike at the station, or using another e-hailing service), which may induce learning about the relative quality of other forms of innovation, thus posing difficulties in the interpretation of results. Secondly, since, as discussed in the introduction, learning about the relative quality of ridesharing vis-à-vis public transportation may have nontrivial policy implications, estimates for this restricted subsample may be also more relevant from this perspective. 

\Cref{balance_main} reports balance tests for the subsample of users close to a train/metro station. We note that groups are well-balanced. There is some evidence at the 10\% level of there being differences in average daily expenses in transportation across groups, though these differences are small when compared to the reported average income. 
% latex table generated in R 4.3.1 by xtable 1.8-4 package
% Tue Oct 24 19:59:40 2023
\begin{table}
\centering
\caption{Balance table: users close to train/metro} 
\label{balance_main}
\scalebox{0.8}{
\begin{tabular}{lllll}
   & Control group & 20\% discount & 50\% discount & p-value no avg diff \\ 
   \hline
Age (years) & 32.5118 & 32.3013 & 32.8516 & 0.6083  \\ 
   & (0.3976) & (0.3852) & (0.3996) &  \\ 
  Female & 0.5531 & 0.5692 & 0.5524 & 0.7244  \\ 
   & (0.0171) & (0.0166) & (0.0171) &  \\ 
  College & 0.7594 & 0.7902 & 0.7503 & 0.1105  \\ 
   & (0.0147) & (0.0136) & (0.0149) &  \\ 
  Monthly income (BRL) & 3469.3396 & 3581.4732 & 3391.0483 & 0.3329  \\ 
   & (91.4375) & (90.9819) & (91.4587) &  \\ 
  Daily expense on transportation (BRL) & 8.9841 & 9.3933 & 8.9257 & 0.0857 * \\ 
   & (0.1632) & (0.1613) & (0.1650) &  \\ 
  Average commuting time (minutes) & 41.3237 & 40.9389 & 42.1401 & 0.6778  \\ 
   & (1.0103) & (0.9324) & (1.0139) &  \\ 
  Regular train/metro user & 0.4564 & 0.4531 & 0.4476 & 0.9347  \\ 
   & (0.0171) & (0.0166) & (0.0171) &  \\ 
     Regular bus user & 0.4564 & 0.4386 & 0.4323 & 0.5854  \\ 
   & (0.0171) & (0.0166) & (0.0170) &  \\ 
  Integrated rides (avg) & 0.0757 & 0.0837 & 0.0870 & 0.6482  \\ 
   & (0.0076) & (0.0079) & (0.0116) &  \\ 
  Non-integrated rides (avg) & 1.3758 & 1.4985 & 1.3434 & 0.2554  \\ 
   & (0.0631) & (0.0718) & (0.0674) &  \\ 
  Integrated rides (t-1) & 0.1097 & 0.1027 & 0.1119 & 0.9173  \\ 
   & (0.0175) & (0.0163) & (0.0163) &  \\ 
  Non-integrated rides (t-1) & 2.0236 & 2.0580 & 1.8634 & 0.2982  \\ 
   & (0.0931) & (0.1028) & (0.0912) &  \\ 
  Integrated rides (t-2) & 0.1167 & 0.1440 & 0.1366 & 0.5143  \\ 
   & (0.0162) & (0.0184) & (0.0234) &  \\ 
  Non-integrated rides (t-2) & 2.1427 & 2.1440 & 1.9623 & 0.3677  \\ 
   & (0.1105) & (0.1074) & (0.1021) &  \\ 
   \hline
N & 848 & 896 & 849 &  \\ 
   \hline
\end{tabular}
}

 \begin{minipage}{0.8\linewidth}
 \vspace{1em}

%       \begin{minipage}{\textwidth} -- what should go in this width argument?
%\begin{singlespace}
            \footnotesize \textit{Notes:} $***$: $p < 0.01$; $**$: $p < 0.05$;  $*$: $p < 0.10$. The first three columns report the average of each trait in the corresponding experimental subgroup. Standard errors are reported in parentheses. The fourth column reports the p-value for a test of the null that the are no average differences across the three groups. Female is an indicator variable of the reported sex being female. College is an indicator variable equal to one if the respondent either goes to or has completed college. Traits accompanied by ``(avg)'' are averages of a biweekly variable in the pretreatment window. Traits accompanied by ``(t - j)'' correspond to a variable $j$ fortnights before the experiment. For reference: 1 BRL = 3.87 USD (December 31st, 2018).
        %    \end{singlespace}
        \end{minipage}

\end{table}

\section{Identification and target estimands}
\label{identification}

\begin{figure}
    \centering
    \includegraphics[scale=0.5]{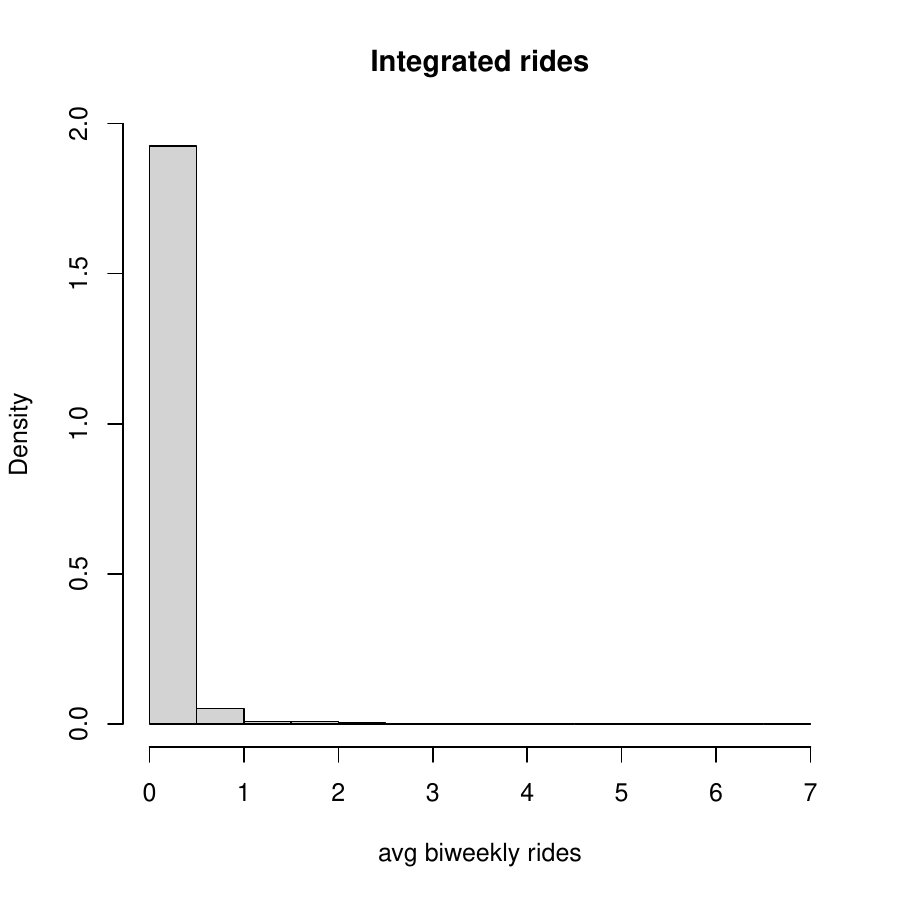}
        \includegraphics[scale=0.5]{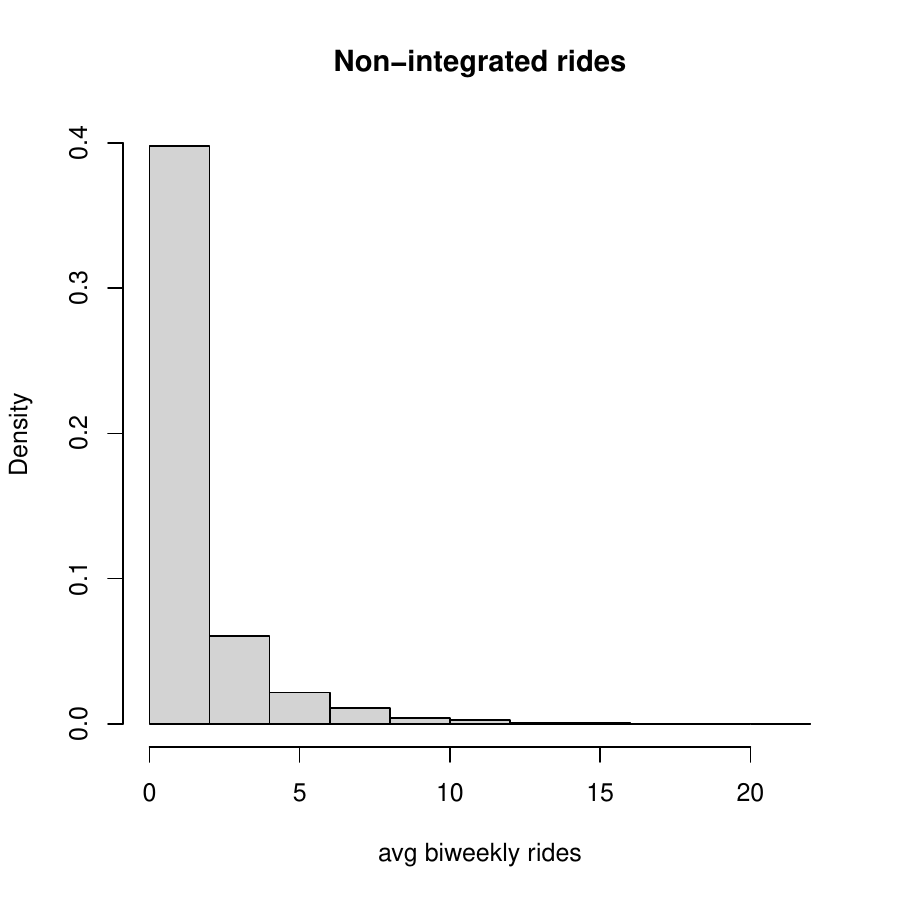}
    \caption{Histograms for pretreatment average biweekly rides}
    \label{fig:hist}
\end{figure}

The distribution of both integrated and non-integrated rides displays heavy-tails. This is evidenced by \Cref{fig:hist}, where we plot histograms for average pre-treatment biweekly rides. The average number of integrated biweekly rides is 0.08, whereas the largest reported value is 6.6. Similarly, the average number of non-integrated rides is 1.4, with the largest reported value being 20.7. 

In settings with heavy tails, difference-in-means estimators of average treatment effects tend to perform poorly, with large sample sizes being required to achieve reasonable precision \citep{Lewis2015,Athey2023}. Following \cite{Athey2023}, we thus opt for semiparametric methods that, while constraining treatment effect heterogeneity, enable the construction of more efficient estimators. 

Guided by our theoretical discussion, we propose a semiparametric model to approximate the effect of discounts on each type of ride. \Cref{prop_learning} shows that the effect of a one-period unanticipated discount in integrated rides has both contemporaneous and extemporaneous effects. Contemporaneously, the magnitude of the impact depends on an agent's information set at the beginning of the period, as well as the technology parameters $\phi$ and $\gamma$. Extemporaneously, the effect is fully mediated by learning. Moreover, our results suggest that, in a world with heterogeneous agents, increased learning may not only change the average number of rides, but also its dispersion. Consequently, a model for treatment effects in our setting should: (a) allow for (some) heterogeneity with respect to prior information sets and the individual technology parameters for bundled goods ($\gamma$ and $\phi$); (b) allow for effects on both the mean and dispersion of outcomes. 

In order to address (a), we consider separate models for respondents who reported being a regular train/metro user, and those that did not. We do so because these individuals may differ in their information regarding $A$ at the beginning of the experiment, as well as in their technology parameters $\gamma$ and $\phi$. We then proceed as follows. Let $b_t(d;u)$ denote the \emph{potential} number of integrated rides at fortnight $t$ for an individual whose prior train/metro usage is $U = u \in \{0,1\}$, upon being assigned treatment $d \in \{0,20\%, 50\%\}$ on the fortnight starting on November 26th, 2018. For each $u \in \{0,1\}$ and $d \in \{20\%, 50\%\}$, we consider the following model:

\begin{equation}
\label{eq_model_b}
    b_t(d;u) = \alpha_{b, t,d,u} + \sigma_{b, t,d,u} b_t(0;u) \, .
\end{equation}

Similarly, definining potential outcomes $n_t(d;u)$ for non-integrated rides, we consider the following model, for $u \in \{0,1\}$ and $d \in \{20\%, 50\%\}$.

\begin{equation}
\label{eq_model_n}
    n_t(d;u) = \alpha_{n, t,d,u} + \sigma_{n, t,d,u} n_t(0;u) \, .
\end{equation}

Consistent with our discussion, models \eqref{eq_model_b} and \eqref{eq_model_n} allow discounts to change both the location and the dispersion of rides. Heterogeneity is unrestricted across prior train/metro usage status $u$ and time $t$, though within an usage-time cell, we restrict treated potential outcomes to follow a location-scale shift with respect to the non-treated potential outcome. Under such restrictions, efficient estimators can be constructed, as we discuss in \Cref{sec_estimation}. 

For the moment, suppose we have estimates $\hat{\alpha}_{b,t,d,u}$ and $\hat{\sigma}_{b, t,d,u}$ from model \eqref{eq_model_b}. In this case, for $d \in \{20\%,50\%\}$, the population average effect in rides $\Delta(t,d,u) \coloneqq \mathbb{E}[b_t(d;u) - b_t(0;u)|U=u]$ may be estimated as:

\begin{equation}\label{eq_estimator_avg}
    \widehat{\Delta(t,d,u)} = \frac{p_{d,u}}{p_{d,u}+p_{0,u}}\left(\bar{b}_{t,d,u} - \left(\frac{\bar{b}_{t,d,u} - \hat{\alpha}_{b,t,d,u}}{\hat{\sigma}_{b, t,d,u} }\right) \right) + \frac{p_{0,u}}{p_{d,u}+p_{0,u}}(\hat{\alpha}_{b,t,d,u} + (\hat{\sigma}_{b, t,d,u} - 1) \bar{b}_{t,0,u})) \, , 
\end{equation}
where $p_{d,u}$ is the fraction of individuals with prior train/metro usage $u$ assigned to group $d$; and $\bar{b}_{t,d,u}$ is the average of integrated rides in fortnight $t$ among the subgroup with prior train/metro usage $u$ and assigned to arm $d$. Estimator \eqref{eq_estimator_avg} relies on the semiparametric model \eqref{eq_model_b} to impute the relevant missing potential outcomes in the subgroup assigned to arms $d$ and $0$.\footnote{It is possible to construct an even more efficient estimator by combining models for different discounts and using these to input the potential outcome for treatment statuses $d$ and $0$ in the subgroup assigned status $d' \notin \{0,d\}$. While this may lead to a more precise estimator, we note that, in the presence of misspecification bias, it may lead to biases in one semiparametric model contaminating treatment effect estimates in another comparison, even if the model for the latter is correctly specified. This is why we opt for ``self-contained'' (involving a single-model) comparisons when estimating effects. See \Cref{sec_estimation} for further discussion.}

Our model also allows us to easily assess the effects of discounts on dispersion. Let $\Psi(t,d,u) \coloneqq \frac{\sqrt{\mathbb{V}[b_t(d;u)|U=u]}- \sqrt{\mathbb{V}[b_t(0;u)|U=u]}}{\sqrt{\mathbb{V}[b_t(0;u)|U=u]}}$ denote the relative change in dispersion induced by discount $d \in \{20\%,50\%\}$. We then estimate $\Psi(t,d,u)$ as:

\begin{equation}\label{eq_estimator_sd}
    \widehat{\Psi(t,d,u)} = \hat{\sigma}_{b, t,d,u}  - 1 \, .
\end{equation}

In the next section, we discuss our approach to estimating $\hat{\alpha}_{b,t,d,u}$ and $\hat{\sigma}_{b, t,d,u}$.

\section{Estimation}
\label{sec_estimation}

Models \eqref{eq_model_b} and \eqref{eq_model_n} are particular parametrisations of the semiparametric models proposed by \cite{Athey2023}. Consider a setting with a binary treatment and potential outcomes $Y(0)$ and $Y(1)$. \cite{Athey2023} analyse models of the form:

\begin{equation}
\label{eq_general_model}
Y(1) = G(Y(0);\theta_0), \quad \theta_0 \in \Theta \subseteq \mathbb{R}^p \, ,
\end{equation}
where $G$ is a known (up to $\theta_0$) function such that $y \mapsto G(y;\theta)$ is increasing, for every $\theta \in \Theta$. Given a random sample from the population and an experimental design that randomly assigns treatment to a subgroup of the sample, \cite{Athey2023} construct $\sqrt{N}$-consistent estimators for the parameter $\theta_0$ that asymptotically achieve the model's efficiency bound \citep{Newey1990,Bickel1993}. Their estimator is constructed by adding a cross-fitted estimate of the model's efficient influence function to a first-step estimator. Estimation of the efficient influence function further requires (cross-fitted) estimates of the densities of $Y(0)$ and $Y(1)$.  

In this paper, we propose an alternative estimator to the semiparametric model \eqref{eq_general_model} with convenient computational and statistical properties. Our estimation approach relies on \textit{L-moments}, a robust alternative to standard moments. For a distribution function $F$ on the real line with quantile function $Q_F$, \citet{Hosking1990} defines the $r$-th L-moment as:
\begin{equation}
    \lambda_r = \int_0^1 Q_F(u) P_{r-1}^*(u)du \, ,
\end{equation}
where $Q_F(u) \coloneqq \inf\{x \in \mathbb{R}:F(x)\geq u\}$, $u \in [0,1]$, is the $u$-quantile of $F$; and $P^*_{r}(u) = \sum_{k=0}^r (-1)^{r-k} \binom{r}{k} \binom{r+k}{k} u^k$ is a shifted Legendre polynomial. L-moments constitute an alternative to standard moments that is less sensitive to outliers. To see this, consider the second L-moment. In this case, \cite{Hosking1990} shows that $\lambda_2 = \frac{1}{2} \mathbb{E}[|Y_1-Y_2|]$, where $Y_1$ and $Y_2$ are independent random variables identically distributed to $F$. In contrast, we can show that the second centered moment is $\mathbb{V}[Y]= \frac{1}{2}  \mathbb{E}[(Y_1-Y_2)^2]$, which is more sensitive to extreme values. L-moments are also known to characterise any distribution with finite first moment \citep[Theorem 1]{Hosking1990}, a property not generally enjoyed by moments \citep[Example 30.2]{Billingsley2012}.

Our proposed estimator is a semiparametric version of the generalised method of L-moments (GMLM) approach discussed in \cite{alvarez2023inference}. Fix $R \in \mathbb{N}$, $R\geq p$, and let $\boldsymbol{P}_R(u) = (P_{0}^*(u), \ldots, P_{r-1}^*(u))'$. We propose to estimate $\theta_0$ as:

\begin{equation}
\label{eq_estimator_text}
    \begin{aligned}
       \hat{\theta} \in \operatorname{argmin}_{\theta \in \Theta} \left\lVert \int_{\underline{p}}^{\overline{p}} (\hat{Q}_{1}(u) -  G(\hat{Q}_{0}(u);\theta)) \boldsymbol{P}_R(u) du \right\rVert_{W_R}^2 \, ,
    \end{aligned}
\end{equation}
where $\hat{Q}_d$ is the empirical quantile function of the outcome in the group assigned treatment status $d \in \{0,1\}$; $0 \leq \underline{p} \leq \overline{p}\leq 1$ are possible trimming constants; and, for a vector $x \in \mathbb{R}^R$, $\lVert x \rVert_{W_R} = \sqrt{x'W_Rx}$, where $W_R$ is a  $R\times R$ weighting matrix. Following the same logic of generalised method of moments estimators, the GMLM \eqref{eq_estimator_text} estimates $\theta_0$ by minimizing a weighted distance between the first $R$ empirical L-moments in the treatment group and the corresponding ``model-implied'' L-moments $\int_{\underline{p}}^{\overline{p}} G(Q_{0}(u); \theta) P^*_{r-1}(u) du$, where the unknown quantile function of the untreated potential outcome is replaced with its empirical counterpart. Our formulation explicitly accommodates for the possibility of trimming ($\underline{p}>0$ or $\underline{p}<1$), which may be employed in the presence of very extreme observations.

In the location-scale models \eqref{eq_model_b} and \eqref{eq_model_n}, estimator \eqref{eq_estimator_text} can be easily computed. Indeed, in these cases, we have that, for any outcome $o \in \{b,n\}$,  treatment $d \in \{20\%, 50\%\}$, prior train/metro usage $u \in \{0,1\}$ and fortnight $t$:

\begin{equation}
\label{eq_est_paper}
    \begin{aligned}
        \begin{pmatrix}
            \hat{\alpha}_{o,t,d,u} \\
            \hat{\sigma}_{o,t,d,u} 
        \end{pmatrix} = \left(\boldsymbol{X}_{o,t,0,u}'W_R \boldsymbol{X}_{0,t,d,u}\right)^{-1}\boldsymbol{X}_{o,t,0,u}'W_r \boldsymbol{y}_{o,t,d,u} \, ,
    \end{aligned}
\end{equation}
where 
\begin{equation}
    \begin{aligned}
        \boldsymbol{y_{o,t,d,u}} = \int_{\underline{p}}^{\overline{p}} \hat{Q}_{o,t,d,u}(u) \boldsymbol{P}_R(u)du \\ 
         \boldsymbol{X_{o,t,d,u}} = \begin{bmatrix}
            \int_{\underline{p}}^{\overline{p}} \boldsymbol{P}_R(u)du & \int_{\underline{p}}^{\overline{p}}  \hat{Q}_{o,t,0,u}(u) \boldsymbol{P}_R(u)du 
         \end{bmatrix}
    \end{aligned} \, ,
\end{equation}
with $\hat{Q}_{o,t,s,u}$ being the empirical quantile function of outcome $o$ at period $t$, in the subgroup assigned treatment status $s$ and with prior train/metro usage $u$. 

In Online \Cref{app_metrics}, we show that, in addition to a convenient computational formulation, our semiparametric GMLM estimator has attractive statistical properties. Online \Cref{app_asymptotic} shows that, in an asymptotic framework where both the number of observations and the number of L-moments $R$ diverge, estimator \eqref{eq_estimator_text} is consistent and asymptotically normal. In this setting, we derive the optimal weighting matrix for L-moments and show how it can be conveniently estimated through a weighted bootstrap procedure. We also show that, in our setting, the minimum of the optimally-weighted objective function \eqref{eq_estimator_text}, when multiplied by sample size, provides researchers with a specification test of overidentifying restrictions (J-test). In Online \Cref{app_efficiency}, we show that, in the single-outcome and binary treatment setting of \cite{Athey2023}, the optimally weighted L-moment estimator without trimming ($\underline{p}=0$ and $\overline{p}=1$) is asymptotically efficient.\footnote{We restrict our attention to the single outcome with binary treatment case because, when there are multiple outcomes or multiple treatments, it may be possible to construct a more efficient estimator by jointly estimating the parameters from different (sub)models. For example, in our empirical application, we note that equations \eqref{eq_model_b} and \eqref{eq_model_n} impose restrictions on the cross-correlation between integrated and non-integrated rides, contemporaneously as well as between different periods. Similarly, \eqref{eq_model_b} and \eqref{eq_model_n} jointly imply that for any individual in any experimental group, the two missing potential outcomes are identified.  Therefore, by jointly estimating all the parameters for a given subgroup $U=u$, it may be possible to produce more efficient estimators. 

We note, however, that a large drawback of this approach is that it would be much more sensitive to misspecification bias, since incorrect specification of the model \eqref{eq_model_b}/\eqref{eq_model_n} for a single tuple (outcome, period, discount, usage) would contaminate estimates for the remaining tuples with same usage. In order to avoid that, we thus separately estimate effects via \eqref{eq_est_paper} for each tuple (outcome, period, discount, usage), and report p-values of the specification test in Online \Cref{app_asymptotic} for the corresponding specification. 

Finally, we note that, if the parametrisation given by \eqref{eq_model_b}/\eqref{eq_model_n} is considered individually for each tuple, then, by the argument in Online Appendix \ref{app_efficiency}, the estimator \eqref{eq_est_paper} is efficient. Put another way,  \eqref{eq_est_paper} is asymptotically efficient for the semiparametric model that assumes its target submodel to be valid, but does not make any hypotheses about the submodels of the remaining tuples.   } Inspired by the Synthetic Controls literature \citep{abadie2021using}, Online Appendix \ref{app_selection} proposes a method to select $R$ and, possibly, the trimming constants $(\underline{p}, \overline{p})$ by relying on pre-treatment data. Finally, Online \Cref{app_montecarlo} shows that our proposed approach compares favourably to existing methods in a Monte Carlo exercise due to \cite{Athey2023}.
\section{Results}
\label{results}
\subsection{Effects on average rides and dispersion} We begin by reporting the effects of discounts on the average number of rides and their dispersion. We estimate the model parameters via \eqref{eq_est_paper}, with weights given by the estimator of the optimal weighting scheme discussed in Online \Cref{app_asymptotic}. The number of L-moments $R$ is selected according to the tuning procedure discussed in Online \Cref{app_selection}. In light of the Monte Carlo exercise in Online \Cref{app_montecarlo} showing limited gains to trimming, we do not allow for it, thus setting $\underline{p}=0<1=\overline{p}$. Parameter estimates are then used to compute the effects of discounts on average rides and dispersion, according to \eqref{eq_estimator_avg} and \eqref{eq_estimator_sd}. Standard errors are computed via the delta-method.

Figures \ref{fig:bimodal_text} and \ref{fig:unimodal_text} plot estimates of the effects on the mean and dipersion, along with 95\% confidence intervals, across the last two pre-treament fortnights (left of vertical dark line) and the post-treatment window, for each discount regime and outcome. We aggregate effects across prior train/metro usage group by averaging  \eqref{eq_estimator_avg} /\eqref{eq_estimator_sd} across $u$, with weights given by the proportion of usage type $u$ in the sample. Consequently, estimates for average effects may be interpreted as estimating the population average treatment effect, whereas for the dispersion one may interpret them as the expected relative change in dispersion across usage groups.

Figure \ref{fig:bimodal_text} reports results for integrated rides. For the 50\% discount, we observe a large and significant effect on average rides on the fortnight where the treatment is in place, and no significant effects thereafter. The magnitude of the contemporaneous effect is large, corresponding to 60\% of the average in the control group at that fortnight. This result contrasts with the usually smaller estimates of own-price elasticities of demand for auto and public transit typically reported in the literature \citep[Section 2.1.5]{small2007economics}.  We also observe a large estimated effect for the 20\% discount, corresponding to 30\% of the control group average, though this effect is not statistically significant. For both discount regimes, there do not appear to be effects on the average nor on the dispersion of integrated rides after the discount is over.

Figure \ref{fig:unimodal_text} reports results for non-integrated rides. Our results show that, between two and three fortnights after the discount is over, we start observing reductions in both the average and the dispersion of integrated rides. These effects persist for over four months after the end of the discount, with reductions in dispersion between 10-20\%. Effects are more pronounced in the 50\% discount regime, though they are also detected in the 20\% discount. By late March and early April, reductions in average rides in the 50\% regime amount to 15\% of the control group mean.

In order to shed more light on the underlying estimates, Tables \ref{table_not_user_bimodal}, \ref{table_user_bimodal}, \ref{table_not_user_unimodal} and \ref{table_user_unimodal} report estimated effects on the mean and dispersion of each outcome, for each discount and prior train/metro usage group. For comparison, we also report estimates of average effects based on difference-in-means estimators, which corrrespond to the efficient estimator in a nonparametric model that does not impose any structure on potential outcomes beyond finite second moments \citep{Newey1990}. Nonparametric estimates of the effects on dispersion are based on the ratio of estimated standard deviations, minus one. For our semiparametric specifications, we also report the p-value of the overidentifying restrictions J-test discussed in Online \Cref{app_asymptotic}.

A few patterns are worth remarking. First, when estimating average effects, the semiparametric methods indeed lead to smaller standard errors than the difference-in-means estimators. Reductions can be as large as 45\%. We also observe gains in precision when estimating effects on the dispersion of outcomes, with an average reduction in standard errors of 16\%. Second, the location-scale parametrisation indeed appears to be adequate, as the  overidentifying restrictions test only rejects the null at the 10\% level for four specifications. In these four cases, inferences based on the nonparametric estimators lead to similar conclusions.\footnote{For completeness, Online Appendix \ref{app_figures} replicates Figures \ref{fig:bimodal_text} and \ref{fig:unimodal_text}, but replacing the semiparametric estimators used in each aggregation with the corresponding nonparametric estimators reported in Tables \ref{table_not_user_bimodal}, \ref{table_user_bimodal}, \ref{table_not_user_unimodal}, and \ref{table_user_unimodal}. Our main results remain unchanged.} Third, we note that the 50\% discount induces contemporaneous increases in integrated rides for both the regular and non-regular train/metro user subpopulations. Moreover, for the 50\% discount, we detect reductions in the mean and dispersion of non-integrated rides in both subpopulations, but especially among non-regular users. In several cases, however, standard errors are large. Finally, we note that, for the 20\% discount, there is some evidence of reductions in the mean and dispersion of integrated rides among the nonregular user subpopulation, in the first fortnights after the discount is over.

\subsection{Decomposing contemporaneous effects} \label{sec_decomposition} Effects reported in the previous subsection are consistent with the following interpretation through the lenses of the model of learning in \Cref{model}. Upon increasing consumption of integrated rides, consumers learn that the relative quality of non-integrated rides was worse than they previously held (they update their beliefs to a more pessimistic value). This leads to a decrease in the average and dispersion of non-integrated rides, as consumers shift away from the innovation. However, since transportation is an essential good, consumers need to increase consumption of the remaining alternatives to satisfy their target. For bundled rides, the impact of relative quality $A$ is dampened by the technology parameter $\phi$; as a consequence, the negative effect introduced by belief updating is offset by the substitution motive, thus leading to the demand for integrated rides remaining approximately unchanged. Do also note that, as a consequence of these patterns, one would expect an increase in the demand for the (unobserved) outside option. The negative effects on non-integrated rides vanish after five months. We can interpret this pattern, \textit{within the model}, through a catch-up in the beliefs in the control group; or, \textit{outside the model}, through an improvement in the underlying quality $A$ or imperfect recall. Disentangling between these competing explanations would require direct measurements of subjects' assessments of relative quality, which we do not possess. We do note, however, that the trajectory of average nonintegrated rides in the control group is compatible with the catching-up mechanism, as we observe a steady decline in the posttreatment window (see \Cref{fig:avg_control} in Online Appendix \Cref{app_figures}).

In order to better understand the workings of learning in our setting, we combine our experimental estimates with Proposition \ref{prop_learning} in order to quantify how much of the contemporaneous increase in integrated rides can be attributed to a \textit{learning} or \textit{experimentation} motive, i.e. to a desire to accumulate information for one's future selves. Recall that Proposition \ref{prop_learning} provides an additive decomposition of the contemporaneous effect of a one-period increase in bimodal rides into a direct price effect and a learning effect. Therefore, if we can reasonably calibrate the direct price effect, we may use our experimental estimates to recover the learning effect.

Proposition  \ref{prop_learning} shows that the direct price effect depends crucially on the beliefs regarding $A$ at the beginning of the period, the technology parameters $\gamma$ and $\phi$, and the sensitivity-to-target parameter $\lambda$. To calibrate these quantities, we combine our experimental dataset with the \textit{2017 Origin and Destination Survey} (ODS). Funded by the São Paulo public metro company, the ODS collects a representative sample of the larger metropolitan area travel flows. In what follows, we describe in steps how we proceed with the calibration.

\vspace{0.5em}

\noindent \textbf{Calibration of prior beliefs regarding $A$:} We assume that individuals hold a common prior regarding the relative quality $A$, which we calibrate using the 2017 ODS data on the quality of ridesharing trips. Specifically, we measure the quality of a trip in terms of travel distance in kilometres, and assume that the prior on $A$ is given by:

$$A = \frac{\xi  \ell}{\bar{x} \bar{l}}, $$
where $(\xi,\ell)$ are independent log-normal random variables, with $\xi$ representing a trip's speed (travel distance over travel time), and $\ell$ representing travel time. The assumption decomposes travel quality onto two components: a measure of ``productivity'' $\xi$, and heterogeneity with respect to duration $\ell$. We calibrate the mean and variance of each component using the ODS data on ridesharing trips that do not integrate with a train/metro. We then normalise this quality by the average speed ($\bar{x}$) and duration ($\bar{l}$) of the outside option, i.e. nonridesharing trips.   

\noindent \textbf{Calibration of $\phi$ and $\gamma$:} We assume that the quality of integrated rides, $\gamma A^\phi$, also admits a decomposition onto log-normals representing speed and duration. We calibrate these random variables from ODS data on ridesharing trips starting or ending at a train/metro. We then choose $\phi$ and $\gamma$ that match the first and second moments of this representation.

    \noindent \textbf{Calibration of $\lambda$:} We allow $\lambda$ to be heterogeneous across individuals in our sample. In keeping with our interpretation of transportation services as an essential good, we assume that, if an individual chooses not to consume transport services, she loses her income. Considering a biweekly window for the model, we assume that:

$$\frac{\lambda_i}{2} \tau_i = \frac{w_i + \frac{w_i}{(1+r)}}{2} \, ,$$
i.e. the individual loses the present value of her monthly income $w_i$ if she does not meet her target. We assume the agent loses her income for two periods to capture the fact that transitions from and to employment are not frictionless \citep{Meghir2015}. We observe $w_i$ for a subset of the users in the experimental sample. We then set the biweekly interest rates $r$ by considering an annual real interest rate of 4\%.  As for the target $\tau_i$, we impute it as follows. In the experimental sample, users were asked their average time spent going to work. We use this information to estimate, in the ODS, the user's expected travel distance. We multiply this value by 20, reflecting that in a fortnight an individual would come and go to work 20 times. We then normalise this quantiy by the productivity of the outside option. Having imputed $\tau_i$ and observing $w_i$ and $r$, we back out $\lambda_i$.

\vspace{0.5em}

Table \ref{table_results_calibration} presents the results of our exercise. We report estimates and standard errors for both the sample average treatment effect on integrated rides,\footnote{The sample average treatment effect for regime $d \in \{20\%,50\%\}$ is defined as $\mathbf{T}_d = \frac{1}{|\mathcal{N}_d \cup \mathcal{N}_0|}\sum_{i \in \mathcal{N}_d \cup \mathcal{N}_0}(b_{\text{November, 26th}}(d;U_i) - b_{\text{November, 26th}}(d;U_i))$, where $\mathcal{N}_s$ is the set of individuals assigned to experimental group $s$ for which income and travel time data is available at the baseline. Subsample average effects are similarly defined. These parameters are estimated using the semiparametric model. Standard errors are computed using the Delta Method \citep[see][]{Athey2023}.} and the percentage of this effect attributable to the learning motive. We note that, for the 50\% discount, around $40\%$ to $50\%$ of the contemporaneous increase in integrated rides can be attributed to an experimentation motive. The estimated share is larger in the subsample of nonregular train/metro users -- precisely the subpopulation for which we detect the largest share of subsequent learning effects (see Tables \ref{table_not_user_bimodal} and \ref{table_not_user_unimodal}). As for the 20\% discount regime, we do not detect significant effects neither on average rides nor on the share attributable to learning. In this case, estimates for the  full sample are considerably more precise than those in subsamples, with the former indicating that the magnitude of the learning effect, if any, is likely to be small.

\section{Concluding remarks}
\label{conclusion}
This paper showed that temporary subsidies to bundles may induce long-term changes in consumption behaviour due to learning about the quality of one of the constituent goods. We introduced a dynamic model where a consumer learns about an innovation on an essential good through repeated consumption. In the model, consumers leverage a one-period discount on a bundle that contains the innovation partly to increase the informational bequest left to her future selves, a mechanism we labeled the learning or experimentation motive for demand.  

We then assessed the predictions of our theory by relying on data from a randomised field experiment conducted by a large ridesharing service in S\~{a}o Paulo. Given the heavy-tailed nature of demand in our setting, and guided by our theoretical discussion, we proposed a semiparametric specificication for treatment effects in our environment, and introduced an efficient estimator for our parametrisation by considering a ``plug-in'' generalised method of L-moments estimator.

Our semiparametric approach enables us to uncover a range of patterns that are broadly consistent with our theory of learning about an essential good. A ten-day discount in app rides integrating with a train/metro station generates persistent negative effects in the mean and dispersion of the demand for non-integrated rides. These effects last for several months, which, given the essentiality nature of transportation services, is interpreted as further evidence of learning. A simple calibration exercise then shows that, for the largest discount regime, around 40\% to 50\% of the contemporaneous increase in integrated rides may be attributed to the experimentation motive.

Insofar as the outside option in our analysis may be interpreted as public transport (PT), our results would indicate that temporary subsidies to modal integration may generate persistent increases in the demand for PT. Indeed, since the  average demand for integrated rides does not decrease with the discount, and that, given the estimates from our analysis, our model predicts an increase in the demand for the outside option after the discount is over; our results would reveal that, by improving the relative assessment on the quality of PT, temporary subsidies on modal integration may lead to persistent increases in PT takeup. From a policy perspective, this suggests that such incentives may be a useful tool in partially offsetting secular declines in PT usage.

Even if the effects of learning are not taken into account, a simple calculation based on the point estimates for the 50\% discount and the decomposition between learning and direct effects reported in Table \ref{table_results_calibration} suggests that the contemporaneous, direct (not driven by learning) own-price-elasticity of the demand for integrated rides to be at least 0.6.\footnote{Using the information in Table \ref{table_results_calibration}, we calculate elasticities as: $$-\frac{(1-\text{share\_due\_to\_learning})\text{sample\_avg\_effect}}{0.5\text{mean\_control}}\, .$$ This calculation hinges on the assumption that all subsidised rides received a 50\% discount. Given that there was a cap of 10BRL in the discounts (which we do not observe, as price data is not available to us), it may thus be seen as a lower bound to the actual elasticity.} This is much larger than values between 0.3 and 0.4, the typical range used as a rule-of-thumb for own-price-elasticities of the demand for PT \citep[Section 4.5.3]{small2007economics}. Public policies aiming at increasing the share of public transit in total trips typically rely on subsidising tariffs. Our calculations suggest that an alternative policy, to integrate e-hailing with public transit through a discount, may induce larger public transport takeup. The increase in takeup could help paying for the large fixed cost in operating PT systems. Last-mile e-hailing rides would complement the system in a potentially more efficient manner, as low-density routes may be better served through ridesharing than through more costly (and polluting) mass transit alternatives.

There are several venues of research which stem from this work. From a policy perspective, it would be interesting to design experiments that directly measure the outside options, so as to properly quantify the PT takeup induced by both the direct effect and learning. From a market design perspective, it would be important to understand what is the optimal pricing policy of a firm when learning-from-bundling is taken into account. Finally, from the econometric viewpoint, it would be relevant to understand more generally the conditions for semiparametric efficiency of ``plug-in'' GMLM estimators, for example by extending arguments from the generalised method of moments literature \citep{Ackeberg2014}.

\begin{figure}
    \centering
     \begin{subfigure}{\textwidth}
     \centering
     \caption{Average}
     \includegraphics[scale=0.45]{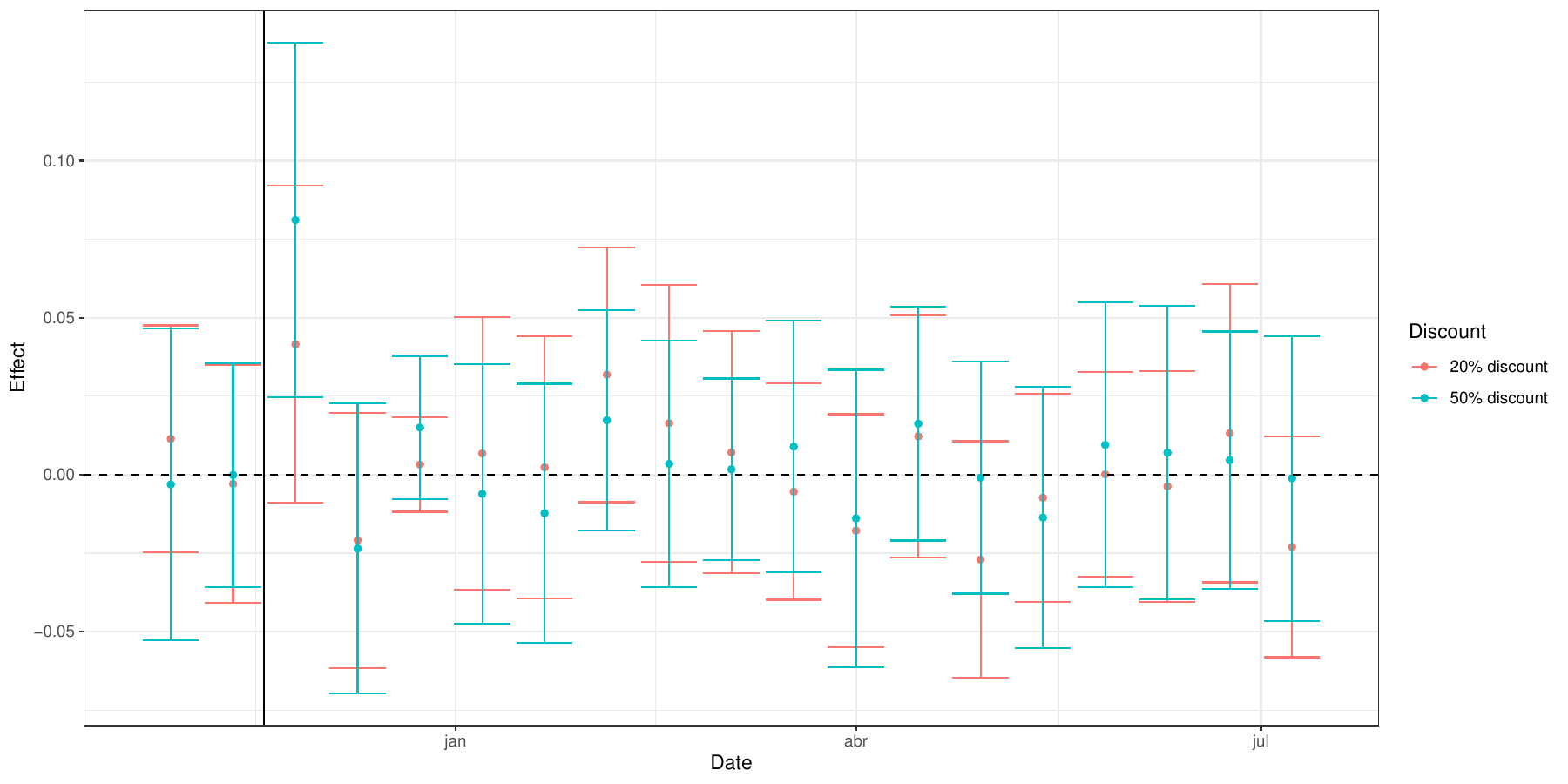}
     \end{subfigure}
          \begin{subfigure}{\textwidth}
     \centering
     \caption{Dispersion}
     \includegraphics[scale=0.45]{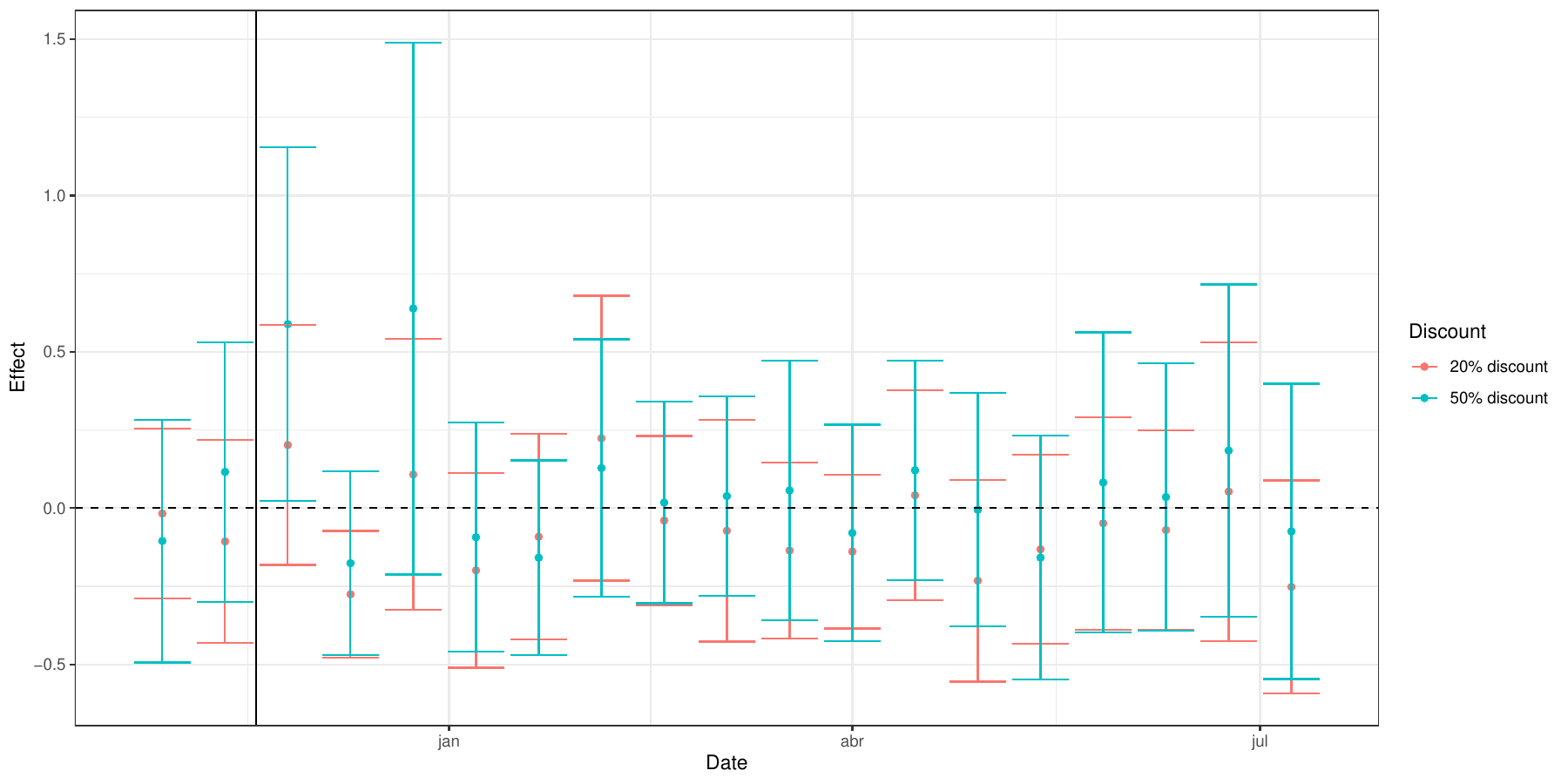}
     \end{subfigure}
    \caption{Effects of discounts on integrated rides}
    \label{fig:bimodal_text}

     \begin{minipage}{1.0\linewidth}
 \vspace{1em}

%       \begin{minipage}{\textwidth} -- what should go in this width argument?
            \footnotesize \textit{Notes:}
        The figure plots estimates, along with 95\% confidence intervals, of the effects of each discount on the mean and dipersion of integrated rides across the last two pre-treament fortnights (left of vertical dark line) and the post-treatment window. We aggregate effects across prior train/metro usage by averaging  \eqref{eq_estimator_avg} /\eqref{eq_estimator_sd} across $u$, with weights given by the proportion of usage type $u$ in the sample. Estimates  \eqref{eq_estimator_avg} /\eqref{eq_estimator_sd} are obtained by first estimating model parameters according to \eqref{eq_est_paper}, with optimal weighting scheme estimated according to the weighted bootstrap in Online Appendix \ref{app_asymptotic}, no trimming, and $R$ selected using pre-treatment data (Online Appendix \ref{app_selection}). Confidence intervals are computed using a normal approximation and the delta method. For reference, the average number of integrated rides in the control (no-discount) group in each fortnight is: \textit{2018-10-29}: 0.12; \textit{2018-11-12}: 0.11; \textit{2018-11-26}: 0.13; \textit{2018-12-10}: 0.14; \textit{2018-12-24}: 0.04; \textit{2019-01-07}: 0.11; \textit{2019-01-21}: 0.11; \textit{2019-02-04}: 0.08; \textit{2019-02-18}: 0.12; \textit{2019-03-04}: 0.1; \textit{2019-03-18}: 0.1; \textit{2019-04-01}: 0.12; \textit{2019-04-15}: 0.11; \textit{2019-04-29}: 0.08; \textit{2019-05-13}: 0.1; \textit{2019-05-27}: 0.09; \textit{2019-06-10}: 0.11; \textit{2019-06-24}: 0.11; \textit{2019-07-08}: 0.1.
    \end{minipage}
    
\end{figure}

\begin{figure}
    \centering
     \begin{subfigure}{\textwidth}
     \centering
     \caption{Average}
     \includegraphics[scale=0.45]{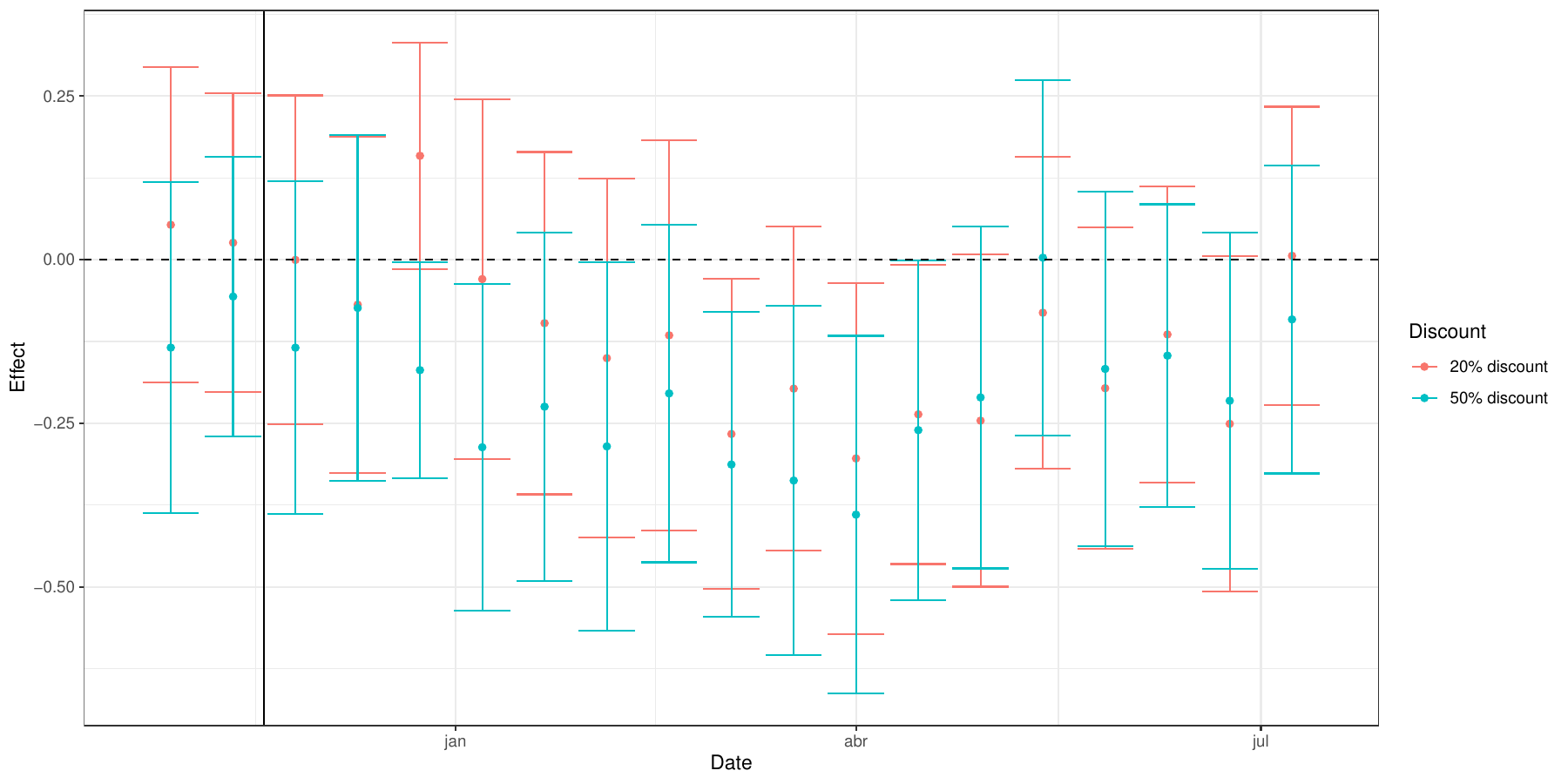}
     \end{subfigure}
          \begin{subfigure}{\textwidth}
     \centering
     \caption{Dispersion}
     \includegraphics[scale=0.45]{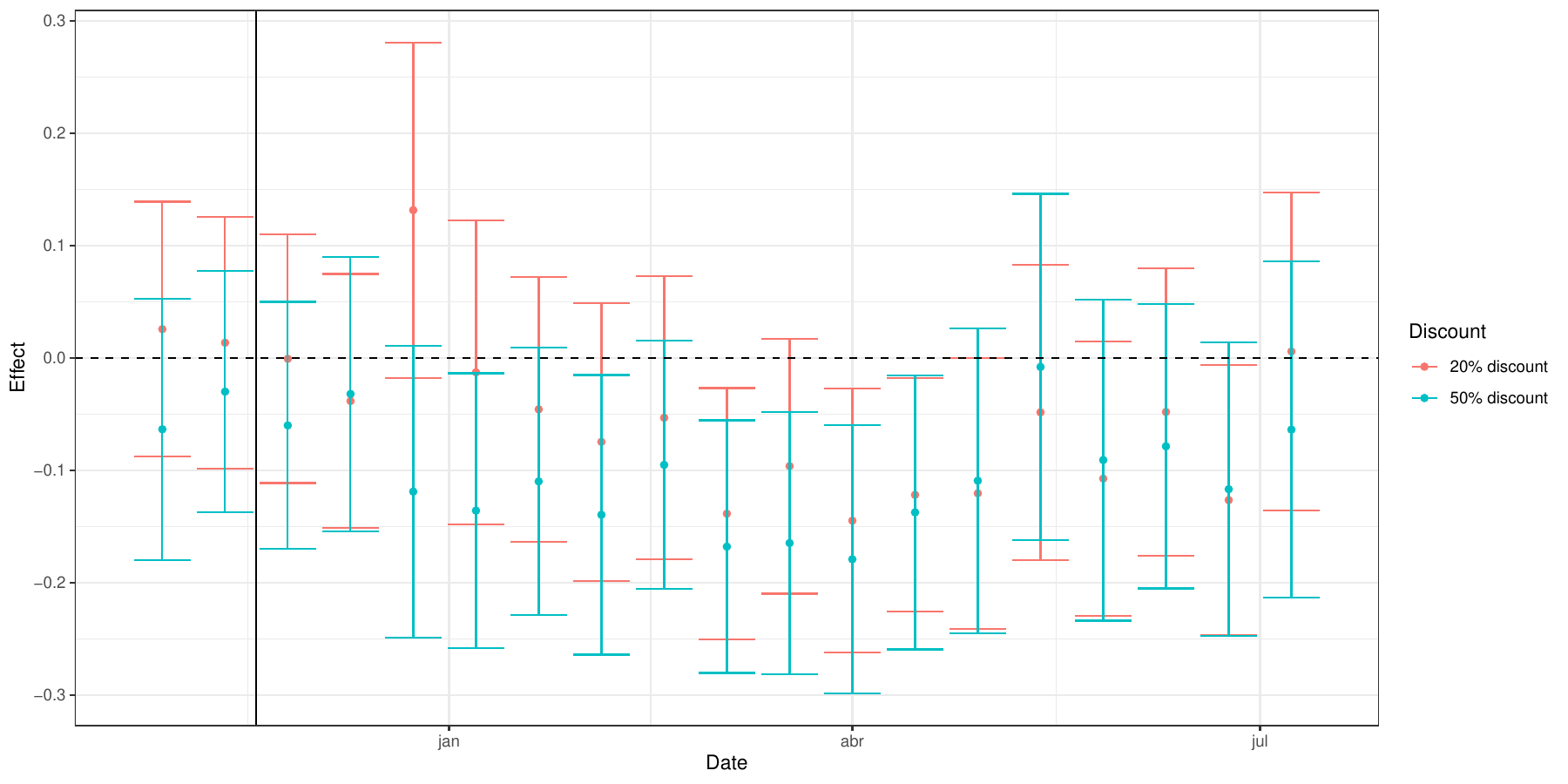}
     \end{subfigure}
    \caption{Effects of discounts on non-integrated rides}
    \label{fig:unimodal_text}

 \begin{minipage}{1.0\linewidth}
 \vspace{1em}

%       \begin{minipage}{\textwidth} -- what should go in this width argument?
            \footnotesize \textit{Notes:}
        The figure plots estimates, along with 95\% confidence intervals, of the effects of each discount on the mean and dipersion of non-integrated rides across the last two pre-treament fortnights (left of vertical dark line) and the post-treatment window. We aggregate effects across prior train/metro usage by averaging  \eqref{eq_estimator_avg} /\eqref{eq_estimator_sd} across $u$, with weights given by the proportion of usage type $u$ in the sample. Estimates  \eqref{eq_estimator_avg} /\eqref{eq_estimator_sd} are obtained by first estimating model parameters according to \eqref{eq_est_paper}, with optimal weighting scheme estimated according to the weighted bootstrap in Online Appendix \ref{app_asymptotic}, no trimming, and $R$ selected using pre-treatment data (Online Appendix \ref{app_selection}). Confidence intervals are computed using a normal approximation and the delta method. For reference, the average number of non-integrated rides in the control (no-discount) group in each fortnight is: \textit{2018-10-29}: 2.14; \textit{2018-11-12}: 2.02; \textit{2018-11-26}: 2.23; \textit{2018-12-10}: 2.15; \textit{2018-12-24}: 1.24; \textit{2019-01-07}: 1.96; \textit{2019-01-21}: 2.11; \textit{2019-02-04}: 2.11; \textit{2019-02-18}: 2.27; \textit{2019-03-04}: 1.9; \textit{2019-03-18}: 2.05; \textit{2019-04-01}: 2.14; \textit{2019-04-15}: 1.97; \textit{2019-04-29}: 1.88; \textit{2019-05-13}: 1.73; \textit{2019-05-27}: 1.84; \textit{2019-06-10}: 1.74; \textit{2019-06-24}: 1.81; \textit{2019-07-08}: 1.52.
    \end{minipage}
\end{figure}

\begin{landscape}
\pagestyle{empty}
    % latex table generated in R 4.3.1 by xtable 1.8-4 package
% Thu Oct 26 21:34:08 2023
\begin{table}[H]
\centering
\caption{Treatment effects on integrated rides: not regular train/metro user} 
\label{table_not_user_bimodal}
\begingroup\footnotesize
\scalebox{0.55}{
\begin{tabular}{lllllllllllllllllll}
  &  & 2018-11-26 & 2018-12-10 & 2018-12-24 & 2019-01-07 & 2019-01-21 & 2019-02-04 & 2019-02-18 & 2019-03-04 & 2019-03-18 & 2019-04-01 & 2019-04-15 & 2019-04-29 & 2019-05-13 & 2019-05-27 & 2019-06-10 & 2019-06-24 & 2019-07-08 \\ 
  \hline \\ & &\multicolumn{ 17 }{c}{\textit{ Integrated rides }}  \\ \cmidrule(lr){3- 19 }\textit{Control group} &  &  &  &  &  &  &  &  &  &  &  &  &  &  &  &  &  &  \\ 
   & Mean & 0.1297 & 0.1427 & 0.0436 & 0.1097 & 0.1132 & 0.0790 & 0.1238 & 0.0979 & 0.0991 & 0.1156 & 0.1085 & 0.0837 & 0.1050 & 0.0920 & 0.1061 & 0.1061 & 0.0979 \\ 
   &  & (0.0035) & (0.0035) & (0.0014) & (0.0030) & (0.0033) & (0.0023) & (0.0028) & (0.0023) & (0.0028) & (0.0029) & (0.0026) & (0.0027) & (0.0026) & (0.0026) & (0.0030) & (0.0028) & (0.0025) \\ 
   &  &  &  &  &  &  &  &  &  &  &  &  &  &  &  &  &  &  \\ 
  \textit{20\% discount} &  &  &  &  &  &  &  &  &  &  &  &  &  &  &  &  &  &  \\ 
   & \textit{Nonparametric model} &  &  &  &  &  &  &  &  &  &  &  &  &  &  &  &  &  \\ 
   &  &  &  &  &  &  &  &  &  &  &  &  &  &  &  &  &  &  \\ 
   & Expected effect (levels) & 0.0454 * & -0.0567 ** & 0.0090  & -0.0595 *** & -0.0482 * & -0.0161  & -0.0282 * & -0.0181  & -0.0096  & -0.0146  & -0.0133  & -0.0093  & -0.0165  & -0.0144  & -0.0259  & 0.0194  & -0.0098  \\ 
   &  & (0.0240)  & (0.0249)  & (0.0115)  & (0.0203)  & (0.0264)  & (0.0198)  & (0.0170)  & (0.0201)  & (0.0206)  & (0.0235)  & (0.0262)  & (0.0216)  & (0.0263)  & (0.0259)  & (0.0303)  & (0.0331)  & (0.0243)  \\ 
   & Effect on dispersion (relative) & 0.1156  & -0.5907 *** & 0.3051  & -0.7178 *** & -0.5553 *** & -0.2066  & -0.3331 ** & 0.1043  & -0.2575  & -0.1507  & 0.0670  & -0.0897  & 0.2299  & -0.1769  & -0.2458  & 1.3757  & 0.3293  \\ 
   &  & (0.2916)  & (0.0968)  & (0.4182)  & (0.0838)  & (0.1567)  & (0.2058)  & (0.1410)  & (0.3338)  & (0.1940)  & (0.1985)  & (0.3410)  & (0.3005)  & (0.5378)  & (0.2980)  & (0.2344)  & (1.1126)  & (0.5382)  \\ 
   &  &  &  &  &  &  &  &  &  &  &  &  &  &  &  &  &  &  \\ 
   & \textit{Semiparametric model (R=14)} &  &  &  &  &  &  &  &  &  &  &  &  &  &  &  &  &  \\ 
   &  &  &  &  &  &  &  &  &  &  &  &  &  &  &  &  &  &  \\ 
   & Expected effect (levels) & 0.0041  & -0.0589 *** & 0.0037  & -0.0431 *** & -0.0407 * & -0.0123  & -0.0162 * & -0.0311  & -0.0213  & -0.0107  & -0.0160  & -0.0098  & -0.0229  & -0.0132  & -0.0279  & -0.0050  & -0.0240  \\ 
   &  & (0.0217)  & (0.0205)  & (0.0076)  & (0.0153)  & (0.0235)  & (0.0140)  & (0.0095)  & (0.0216)  & (0.0159)  & (0.0144)  & (0.0188)  & (0.0159)  & (0.0197)  & (0.0214)  & (0.0249)  & (0.0253)  & (0.0229)  \\ 
   & Effect on dispersion (relative) & 0.0504  & -0.6067 *** & 0.1663  & -0.7078 *** & -0.4734 *** & -0.1955  & -0.2914 ** & -0.4230 * & -0.3235 * & -0.1496  & -0.1815  & -0.1798  & -0.3038  & -0.1857  & -0.2957  & -0.0762  & -0.3354  \\ 
   &  & (0.2768)  & (0.0895)  & (0.3711)  & (0.0899)  & (0.1694)  & (0.1983)  & (0.1397)  & (0.2466)  & (0.1782)  & (0.1842)  & (0.1977)  & (0.2674)  & (0.2298)  & (0.2680)  & (0.2153)  & (0.3808)  & (0.2818)  \\ 
   & p-value overid test & 0.6298  & 0.9982  & 1.0000  & 0.9936  & 0.9989  & 1.0000  & 1.0000  & 0.9141  & 0.9999  & 1.0000  & 1.0000  & 1.0000  & 1.0000  & 1.0000  & 1.0000  & 0.9999  & 0.9988  \\ 
   &  &  &  &  &  &  &  &  &  &  &  &  &  &  &  &  &  &  \\ 
  \textit{50\% discount} &  &  &  &  &  &  &  &  &  &  &  &  &  &  &  &  &  &  \\ 
   & \textit{Nonparametric model} &  &  &  &  &  &  &  &  &  &  &  &  &  &  &  &  &  \\ 
   &  &  &  &  &  &  &  &  &  &  &  &  &  &  &  &  &  &  \\ 
   & Expected effect (levels) & 0.1120 *** & -0.0166  & 0.0338 * & -0.0119  & -0.0378  & -0.0118  & 0.0053  & -0.0054  & 0.0075  & -0.0013  & -0.0121  & -0.0009  & -0.0140  & 0.0030  & -0.0144  & 0.0354  & 0.0032  \\ 
   &  & (0.0341)  & (0.0303)  & (0.0185)  & (0.0276)  & (0.0270)  & (0.0200)  & (0.0224)  & (0.0200)  & (0.0244)  & (0.0261)  & (0.0259)  & (0.0229)  & (0.0263)  & (0.0290)  & (0.0334)  & (0.0288)  & (0.0289)  \\ 
   & Effect on dispersion (relative) & 0.8546 * & -0.1478  & 1.4044  & 0.0018  & -0.4993 *** & -0.2026  & 0.2241  & 0.0663  & 0.0680  & 0.0472  & 0.0182  & 0.0093  & 0.2041  & 0.0348  & -0.0603  & 0.9585  & 0.6768  \\ 
   &  & (0.5030)  & (0.2605)  & (0.8587)  & (0.3836)  & (0.1803)  & (0.1939)  & (0.3768)  & (0.2389)  & (0.3207)  & (0.2722)  & (0.2294)  & (0.2866)  & (0.4075)  & (0.3306)  & (0.3228)  & (0.7439)  & (0.5837)  \\ 
   &  &  &  &  &  &  &  &  &  &  &  &  &  &  &  &  &  &  \\ 
   & \textit{Semiparametric model (R=4)} &  &  &  &  &  &  &  &  &  &  &  &  &  &  &  &  &  \\ 
   &  &  &  &  &  &  &  &  &  &  &  &  &  &  &  &  &  &  \\ 
   & Expected effect (levels) & 0.0551 ** & -0.0206  & 0.0253 * & -0.0142  & -0.0276  & -0.0117  & 0.0001  & 0.0050  & -0.0007  & -0.0018  & 0.0022  & 0.0003  & -0.0177  & 0.0020  & -0.0137  & 0.0244  & -0.0204  \\ 
   &  & (0.0275)  & (0.0246)  & (0.0142)  & (0.0228)  & (0.0209)  & (0.0140)  & (0.0165)  & (0.0150)  & (0.0199)  & (0.0195)  & (0.0184)  & (0.0138)  & (0.0241)  & (0.0245)  & (0.0280)  & (0.0234)  & (0.0328)  \\ 
   & Effect on dispersion (relative) & 0.6696  & -0.2108  & 1.1348  & -0.1894  & -0.3246 * & -0.1795  & 0.0017  & 0.0838  & -0.0107  & -0.0241  & 0.0271  & 0.0062  & -0.2406  & 0.0273  & -0.1476  & 0.4302  & -0.2665  \\ 
   &  & (0.4610)  & (0.2178)  & (0.7596)  & (0.2781)  & (0.1865)  & (0.1898)  & (0.2514)  & (0.2545)  & (0.3147)  & (0.2610)  & (0.2333)  & (0.2579)  & (0.3039)  & (0.3440)  & (0.2808)  & (0.4659)  & (0.3897)  \\ 
   & p-value overid test & 0.0313 ** & 0.9612  & 0.8306  & 0.8736  & 0.3586  & 0.8112  & 0.8920  & 0.2650  & 0.8661  & 0.8532  & 0.3386  & 0.9459  & 0.2523  & 0.9977  & 0.9848  & 0.8026  & 0.1043  \\ 
   \hline
\end{tabular}
}
\begin{minipage}{0.9\linewidth}
 \vspace{1em}

%       \begin{minipage}{\textwidth} -- what should go in this width argument?
            \tiny \textit{Notes:} $***$: $p < 0.01$; $**$: $p < 0.05$;  $*$: $p < 0.10$. Standard error in parentheses. Sample sizes for each experimental arm are:  461 (control), 490 (20\% discount) and 469 (50\% discount). Estimates for expected effects under ``Nonparametric model'' correspond to difference-in-means estimators, which are efficient when no restrictions on potential outcomes are imposed \citep{Newey1990}. Estimates for effects on dispersion under ``Nonparametric model'' are based on the ratio of estimated standard deviations minus one, with standard errors computed via the delta method. Estimates under ``Semiparametric model'' are constructed using model \eqref{eq_model_b} and estimator \eqref{eq_estimator_text}, with optimal weighting scheme estimated according to the weighted bootstrap in Online Appendix \ref{app_asymptotic}, no trimming, and $R$ selected using pre-treatment data (Online Appendix \ref{app_selection}). The expected effect in the semiparametric model is computed using \eqref{eq_estimator_avg}. The effect on dispersion is computed using \eqref{eq_estimator_sd}. Standard errors in the semiparametric model are constructed via the delta method. Line ``p-value overid test'' reports the p-value for the specification test discussed in Online \Cref{app_asymptotic}
            \end{minipage}
\endgroup
\end{table}

    % latex table generated in R 4.3.1 by xtable 1.8-4 package
% Thu Oct 26 21:34:08 2023
\begin{table}[H]
\centering
\caption{Treatment effects on integrated rides: regular train/metro user} 
\label{table_user_bimodal}
\begingroup\footnotesize
\scalebox{0.55}{
\begin{tabular}{lllllllllllllllllll}
  &  & 2018-11-26 & 2018-12-10 & 2018-12-24 & 2019-01-07 & 2019-01-21 & 2019-02-04 & 2019-02-18 & 2019-03-04 & 2019-03-18 & 2019-04-01 & 2019-04-15 & 2019-04-29 & 2019-05-13 & 2019-05-27 & 2019-06-10 & 2019-06-24 & 2019-07-08 \\ 
  \hline \\ & &\multicolumn{ 17 }{c}{\textit{ Integrated rides }}  \\ \cmidrule(lr){3- 19 }\textit{Control group} &  &  &  &  &  &  &  &  &  &  &  &  &  &  &  &  &  &  \\ 
   & Mean & 0.1297 & 0.1427 & 0.0436 & 0.1097 & 0.1132 & 0.0790 & 0.1238 & 0.0979 & 0.0991 & 0.1156 & 0.1085 & 0.0837 & 0.1050 & 0.0920 & 0.1061 & 0.1061 & 0.0979 \\ 
   &  & (0.0035) & (0.0035) & (0.0014) & (0.0030) & (0.0033) & (0.0023) & (0.0028) & (0.0023) & (0.0028) & (0.0029) & (0.0026) & (0.0027) & (0.0026) & (0.0026) & (0.0030) & (0.0028) & (0.0025) \\ 
   &  &  &  &  &  &  &  &  &  &  &  &  &  &  &  &  &  &  \\ 
  \textit{20\% discount} &  &  &  &  &  &  &  &  &  &  &  &  &  &  &  &  &  &  \\ 
   & \textit{Nonparametric model} &  &  &  &  &  &  &  &  &  &  &  &  &  &  &  &  &  \\ 
   &  &  &  &  &  &  &  &  &  &  &  &  &  &  &  &  &  &  \\ 
   & Expected effect (levels) & 0.1182 * & 0.0326  & 0.0040  & 0.0889  & 0.0723  & 0.1408 ** & 0.0893  & 0.0872 * & 0.0227  & 0.0538  & 0.0799 * & 0.0240  & 0.0252  & 0.0611  & 0.0362  & 0.0290  & -0.0164  \\ 
   &  & (0.0657)  & (0.0543)  & (0.0226)  & (0.0570)  & (0.0492)  & (0.0583)  & (0.0570)  & (0.0451)  & (0.0440)  & (0.0440)  & (0.0447)  & (0.0382)  & (0.0399)  & (0.0384)  & (0.0394)  & (0.0515)  & (0.0379)  \\ 
   & Effect on dispersion (relative) & 0.2925  & 0.1574  & 0.0874  & 0.5156  & 0.1601  & 1.2780  & 0.4797  & 0.4865  & 0.0661  & 0.0636  & 0.3497  & -0.1791  & 0.1250  & 0.2111  & 0.1443  & 0.1903  & -0.0606  \\ 
   &  & (0.3169)  & (0.2502)  & (0.2129)  & (0.4494)  & (0.3435)  & (0.8967)  & (0.2948)  & (0.3058)  & (0.2722)  & (0.1844)  & (0.3130)  & (0.1942)  & (0.2079)  & (0.2316)  & (0.3292)  & (0.2777)  & (0.2668)  \\ 
   &  &  &  &  &  &  &  &  &  &  &  &  &  &  &  &  &  &  \\ 
   & \textit{Semiparametric model (R=14)} &  &  &  &  &  &  &  &  &  &  &  &  &  &  &  &  &  \\ 
   &  &  &  &  &  &  &  &  &  &  &  &  &  &  &  &  &  &  \\ 
   & Expected effect (levels) & 0.0865 * & 0.0246  & 0.0027  & 0.0666  & 0.0540  & 0.0848 ** & 0.0555  & 0.0529  & 0.0136  & -0.0264  & 0.0460  & -0.0477  & 0.0114  & 0.0160  & 0.0253  & 0.0350  & -0.0218  \\ 
   &  & (0.0504)  & (0.0384)  & (0.0142)  & (0.0451)  & (0.0375)  & (0.0424)  & (0.0482)  & (0.0346)  & (0.0336)  & (0.0379)  & (0.0370)  & (0.0378)  & (0.0288)  & (0.0262)  & (0.0284)  & (0.0438)  & (0.0283)  \\ 
   & Effect on dispersion (relative) & 0.3836  & 0.1214  & 0.0372  & 0.4111  & 0.3661  & 0.7263  & 0.2621  & 0.3479  & 0.0895  & -0.1261  & 0.3081  & -0.2950 * & 0.0749  & 0.1160  & 0.2002  & 0.2080  & -0.1523  \\ 
   &  & (0.2744)  & (0.2001)  & (0.1970)  & (0.3324)  & (0.3077)  & (0.4527)  & (0.2531)  & (0.2659)  & (0.2335)  & (0.1657)  & (0.2917)  & (0.1690)  & (0.1973)  & (0.2056)  & (0.2483)  & (0.2823)  & (0.1794)  \\ 
   & p-value overid test & 0.9998  & 1.0000  & 1.0000  & 1.0000  & 0.9998  & 0.9997  & 0.6041  & 0.9997  & 1.0000  & 0.5258  & 0.9974  & 0.5819  & 0.9997  & 0.9936  & 1.0000  & 1.0000  & 1.0000  \\ 
   &  &  &  &  &  &  &  &  &  &  &  &  &  &  &  &  &  &  \\ 
  \textit{50\% discount} &  &  &  &  &  &  &  &  &  &  &  &  &  &  &  &  &  &  \\ 
   & \textit{Nonparametric model} &  &  &  &  &  &  &  &  &  &  &  &  &  &  &  &  &  \\ 
   &  &  &  &  &  &  &  &  &  &  &  &  &  &  &  &  &  &  \\ 
   & Expected effect (levels) & 0.1592 ** & -0.0227  & -0.0013  & 0.0186  & 0.0236  & 0.0754  & 0.0194  & 0.0104  & 0.0238  & 0.0267  & 0.0498  & 0.0259  & 0.0000  & 0.0390  & 0.0391  & -0.0205  & 0.0236  \\ 
   &  & (0.0662)  & (0.0518)  & (0.0247)  & (0.0461)  & (0.0491)  & (0.0470)  & (0.0501)  & (0.0367)  & (0.0480)  & (0.0545)  & (0.0444)  & (0.0416)  & (0.0440)  & (0.0476)  & (0.0459)  & (0.0447)  & (0.0422)  \\ 
   & Effect on dispersion (relative) & 0.2651  & 0.0234  & 0.2259  & 0.0350  & 0.1163  & 0.6766  & 0.1677  & 0.0195  & 0.2058  & 0.4630  & 0.2936  & -0.0356  & 0.2854  & 0.6247  & 0.4155  & -0.1333  & 0.1217  \\ 
   &  & (0.3346)  & (0.2580)  & (0.3308)  & (0.2869)  & (0.3821)  & (0.6023)  & (0.2361)  & (0.2158)  & (0.3554)  & (0.3733)  & (0.3172)  & (0.2874)  & (0.3479)  & (0.4711)  & (0.4373)  & (0.1940)  & (0.2461)  \\ 
   &  &  &  &  &  &  &  &  &  &  &  &  &  &  &  &  &  &  \\ 
   & \textit{Semiparametric model (R=7)} &  &  &  &  &  &  &  &  &  &  &  &  &  &  &  &  &  \\ 
   &  &  &  &  &  &  &  &  &  &  &  &  &  &  &  &  &  &  \\ 
   & Expected effect (levels) & 0.1127 ** & -0.0270  & 0.0026  & 0.0037  & 0.0063  & 0.0525  & 0.0075  & -0.0024  & 0.0205  & -0.0286  & 0.0333  & -0.0025  & -0.0087  & 0.0186  & 0.0321  & -0.0194  & 0.0222  \\ 
   &  & (0.0544)  & (0.0428)  & (0.0191)  & (0.0377)  & (0.0390)  & (0.0359)  & (0.0395)  & (0.0272)  & (0.0383)  & (0.0479)  & (0.0355)  & (0.0382)  & (0.0368)  & (0.0417)  & (0.0404)  & (0.0366)  & (0.0323)  \\ 
   & Effect on dispersion (relative) & 0.4906  & -0.1342  & 0.0373  & 0.0232  & 0.0426  & 0.5011  & 0.0371  & -0.0165  & 0.1381  & -0.1466  & 0.2351  & -0.0188  & -0.0586  & 0.1484  & 0.2577  & -0.1140  & 0.1584  \\ 
   &  & (0.3079)  & (0.1998)  & (0.2735)  & (0.2397)  & (0.2693)  & (0.4041)  & (0.1985)  & (0.1876)  & (0.2708)  & (0.2295)  & (0.2782)  & (0.2835)  & (0.2427)  & (0.3465)  & (0.3424)  & (0.2017)  & (0.2471)  \\ 
   & p-value overid test & 0.2361  & 0.9912  & 0.8387  & 0.9866  & 0.9983  & 0.9936  & 0.9812  & 0.9976  & 0.9999  & 0.5774  & 0.9969  & 0.6037  & 0.7805  & 0.8223  & 0.9603  & 0.9998  & 1.0000  \\ 
   \hline
\end{tabular}
}
\endgroup
\begin{minipage}{0.9\linewidth}
 \vspace{1em}
 
%       \begin{minipage}{\textwidth} -- what should go in this width argument?
            \tiny \textit{Notes:} $***$: $p < 0.01$; $**$: $p < 0.05$;  $*$: $p < 0.10$. Standard error in parentheses. Sample sizes for each experimental arm are:  387 (control), 406 (20\% discount) and 380 (50\% discount). Estimates for expected effects under ``Nonparametric model'' correspond to difference-in-means estimators, which are efficient when no restrictions on potential outcomes are imposed \citep{Newey1990}. Estimates for effects on dispersion under ``Nonparametric model'' are based on the ratio of estimated standard deviations minus one, with standard errors computed via the delta method. Estimates under ``Semiparametric model'' are constructed using model \eqref{eq_model_b} and estimator \eqref{eq_estimator_text}, with optimal weighting scheme estimated according to the weighted bootstrap in Online Appendix \ref{app_asymptotic}, no trimming, and $R$ selected using pre-treatment data (Online Appendix \ref{app_selection}). The expected effect in the semiparametric model is computed using \eqref{eq_estimator_avg}. The effect on dispersion is computed using \eqref{eq_estimator_sd}. Standard errors in the semiparametric model are constructed via the delta method. Line ``p-value overid test'' reports the p-value for the specification test discussed in Online \Cref{app_asymptotic}
            \end{minipage}
\end{table}

    % latex table generated in R 4.3.1 by xtable 1.8-4 package
% Thu Oct 26 21:34:08 2023
\begin{table}[H]
\centering
\caption{Treatment effects on non-integrated rides: not regular train/metro user} 
\label{table_not_user_unimodal}
\begingroup\footnotesize
\scalebox{0.55}{
\begin{tabular}{lllllllllllllllllll}
  &  & 2018-11-26 & 2018-12-10 & 2018-12-24 & 2019-01-07 & 2019-01-21 & 2019-02-04 & 2019-02-18 & 2019-03-04 & 2019-03-18 & 2019-04-01 & 2019-04-15 & 2019-04-29 & 2019-05-13 & 2019-05-27 & 2019-06-10 & 2019-06-24 & 2019-07-08 \\ 
  \hline \\ & &\multicolumn{ 17 }{c}{\textit{ Non-integrated rides }}  \\ \cmidrule(lr){3- 19 }\textit{Control group} &  &  &  &  &  &  &  &  &  &  &  &  &  &  &  &  &  &  \\ 
   & Mean & 2.2252 & 2.1450 & 1.2417 & 1.9587 & 2.1108 & 2.1144 & 2.2653 & 1.9045 & 2.0472 & 2.1356 & 1.9741 & 1.8762 & 1.7252 & 1.8361 & 1.7358 & 1.8137 & 1.5189 \\ 
   &  & (0.0185) & (0.0190) & (0.0125) & (0.0188) & (0.0191) & (0.0199) & (0.0203) & (0.0179) & (0.0198) & (0.0206) & (0.0191) & (0.0195) & (0.0183) & (0.0192) & (0.0176) & (0.0188) & (0.0163) \\ 
   &  &  &  &  &  &  &  &  &  &  &  &  &  &  &  &  &  &  \\ 
  \textit{20\% discount} &  &  &  &  &  &  &  &  &  &  &  &  &  &  &  &  &  &  \\ 
   & \textit{Nonparametric model} &  &  &  &  &  &  &  &  &  &  &  &  &  &  &  &  &  \\ 
   &  &  &  &  &  &  &  &  &  &  &  &  &  &  &  &  &  &  \\ 
   & Expected effect (levels) & -0.0063  & -0.0070  & 0.0451  & -0.0668  & -0.0482  & -0.2077  & -0.0703  & -0.2824  & -0.2524  & -0.4893 ** & -0.1041  & -0.1434  & -0.0538  & -0.2491  & -0.1747  & 0.0770  & 0.1195  \\ 
   &  & (0.2261)  & (0.2465)  & (0.1557)  & (0.2421)  & (0.2391)  & (0.2399)  & (0.2486)  & (0.2105)  & (0.2280)  & (0.2418)  & (0.2176)  & (0.2220)  & (0.2106)  & (0.2223)  & (0.2091)  & (0.2134)  & (0.1960)  \\ 
   & Effect on dispersion (relative) & -0.0100  & 0.0388  & 0.0452  & 0.0509  & 0.0581  & -0.0056  & 0.0227  & -0.0538  & 0.0229  & -0.0668  & -0.0623  & -0.1187  & -0.0658  & -0.1267  & -0.1220  & -0.0799  & 0.0551  \\ 
   &  & (0.1123)  & (0.1134)  & (0.1317)  & (0.1147)  & (0.1331)  & (0.1287)  & (0.1345)  & (0.1144)  & (0.1080)  & (0.1084)  & (0.0981)  & (0.0847)  & (0.0946)  & (0.1067)  & (0.0899)  & (0.0929)  & (0.1270)  \\ 
   &  &  &  &  &  &  &  &  &  &  &  &  &  &  &  &  &  &  \\ 
   & \textit{Semiparametric model (R=8)} &  &  &  &  &  &  &  &  &  &  &  &  &  &  &  &  &  \\ 
   &  &  &  &  &  &  &  &  &  &  &  &  &  &  &  &  &  &  \\ 
   & Expected effect (levels) & 0.0124  & 0.0113  & 0.1622  & -0.0568  & -0.1011  & -0.1229  & -0.0233  & -0.2866  & -0.1974  & -0.3246  & -0.1158  & -0.3166 * & -0.0477  & -0.1809  & -0.2926 * & -0.2689  & -0.0420  \\ 
   &  & (0.1891)  & (0.1981)  & (0.1328)  & (0.2150)  & (0.2024)  & (0.2132)  & (0.2278)  & (0.1794)  & (0.1842)  & (0.2026)  & (0.1804)  & (0.1759)  & (0.1748)  & (0.1884)  & (0.1653)  & (0.1752)  & (0.1663)  \\ 
   & Effect on dispersion (relative) & 0.0051  & 0.0046  & 0.1228  & -0.0246  & -0.0426  & -0.0533  & -0.0098  & -0.1375 * & -0.0911  & -0.1403 * & -0.0541  & -0.1431 ** & -0.0253  & -0.0906  & -0.1449 * & -0.1301 * & -0.0243  \\ 
   &  & (0.0788)  & (0.0798)  & (0.1057)  & (0.0926)  & (0.0840)  & (0.0894)  & (0.0955)  & (0.0800)  & (0.0814)  & (0.0825)  & (0.0815)  & (0.0729)  & (0.0905)  & (0.0893)  & (0.0746)  & (0.0782)  & (0.0973)  \\ 
   & p-value overid test & 0.9726  & 0.9956  & 0.2687  & 0.5043  & 0.9597  & 0.7129  & 0.9946  & 0.8701  & 0.4621  & 0.2400  & 0.9871  & 0.8120  & 0.5507  & 0.9289  & 0.9904  & 0.2770  & 0.4572  \\ 
   &  &  &  &  &  &  &  &  &  &  &  &  &  &  &  &  &  &  \\ 
  \textit{50\% discount} &  &  &  &  &  &  &  &  &  &  &  &  &  &  &  &  &  &  \\ 
   & \textit{Nonparametric model} &  &  &  &  &  &  &  &  &  &  &  &  &  &  &  &  &  \\ 
   &  &  &  &  &  &  &  &  &  &  &  &  &  &  &  &  &  &  \\ 
   & Expected effect (levels) & -0.0668  & -0.2004  & -0.2580 * & -0.5616 *** & -0.3042  & -0.4157 * & -0.2477  & -0.3532 * & -0.3360  & -0.5016 ** & -0.2625  & -0.2834  & 0.0634  & -0.3162  & -0.1314  & -0.0575  & -0.0294  \\ 
   &  & (0.2295)  & (0.2506)  & (0.1418)  & (0.2147)  & (0.2315)  & (0.2386)  & (0.2428)  & (0.2052)  & (0.2215)  & (0.2283)  & (0.2112)  & (0.2265)  & (0.2263)  & (0.2244)  & (0.2129)  & (0.2191)  & (0.1879)  \\ 
   & Effect on dispersion (relative) & -0.0017  & 0.0494  & -0.1670 * & -0.2162 ** & -0.0296  & -0.0381  & -0.0476  & -0.1260  & -0.0574  & -0.2041 ** & -0.1426  & -0.0969  & 0.0588  & -0.1266  & -0.1042  & -0.0466  & -0.0529  \\ 
   &  & (0.1063)  & (0.1263)  & (0.0965)  & (0.0847)  & (0.0966)  & (0.1092)  & (0.1093)  & (0.0981)  & (0.0944)  & (0.0855)  & (0.0879)  & (0.0951)  & (0.1164)  & (0.1112)  & (0.0956)  & (0.1055)  & (0.1156)  \\ 
   &  &  &  &  &  &  &  &  &  &  &  &  &  &  &  &  &  &  \\ 
   & \textit{Semiparametric model (R=8)} &  &  &  &  &  &  &  &  &  &  &  &  &  &  &  &  &  \\ 
   &  &  &  &  &  &  &  &  &  &  &  &  &  &  &  &  &  &  \\ 
   & Expected effect (levels) & -0.1321  & -0.1289  & -0.3147 *** & -0.4543 ** & -0.1723  & -0.2822  & -0.0957  & -0.3246 * & -0.2936 * & -0.4917 ** & -0.2094  & -0.2800  & 0.1608  & -0.2115  & -0.2056  & -0.2164  & 0.0173  \\ 
   &  & (0.1802)  & (0.1964)  & (0.1068)  & (0.1862)  & (0.1957)  & (0.2089)  & (0.1788)  & (0.1670)  & (0.1745)  & (0.1920)  & (0.1824)  & (0.1838)  & (0.2012)  & (0.1973)  & (0.1662)  & (0.1724)  & (0.1594)  \\ 
   & Effect on dispersion (relative) & -0.0539  & -0.0524  & -0.2239 *** & -0.2029 *** & -0.0761  & -0.1232  & -0.0413  & -0.1571 ** & -0.1349 * & -0.2046 *** & -0.0994  & -0.1325  & 0.0860  & -0.1071  & -0.1031  & -0.1107  & 0.0111  \\ 
   &  & (0.0712)  & (0.0781)  & (0.0658)  & (0.0740)  & (0.0836)  & (0.0855)  & (0.0760)  & (0.0734)  & (0.0744)  & (0.0712)  & (0.0817)  & (0.0818)  & (0.1119)  & (0.0941)  & (0.0784)  & (0.0828)  & (0.1016)  \\ 
   & p-value overid test & 0.9289  & 0.3837  & 0.3264  & 0.3726  & 0.6104  & 0.4401  & 0.2445  & 0.7636  & 0.4456  & 0.7125  & 0.5503  & 0.6254  & 0.2253  & 0.8497  & 0.9878  & 0.9606  & 0.6405  \\ 
   \hline
\end{tabular}
}

\begin{minipage}{0.9\linewidth}
 \vspace{1em}
%       \begin{minipage}{\textwidth} -- what should go in this width argument?
            \tiny \textit{Notes:} $***$: $p < 0.01$; $**$: $p < 0.05$;  $*$: $p < 0.10$. Standard error in parentheses. Sample sizes for each experimental arm are:  461 (control), 490 (20\% discount) and 469 (50\% discount). Estimates for expected effects under ``Nonparametric model'' correspond to difference-in-means estimators, which are efficient when no restrictions on potential outcomes are imposed \citep{Newey1990}. Estimates for effects on dispersion under ``Nonparametric model'' are based on the ratio of estimated standard deviations minus one, with standard errors computed via the delta method. Estimates under ``Semiparametric model'' are constructed using model \eqref{eq_model_n} and estimator \eqref{eq_estimator_text}, with optimal weighting scheme estimated according to the weighted bootstrap in Online Appendix \ref{app_asymptotic}, no trimming, and $R$ selected using pre-treatment data (Online Appendix \ref{app_selection}). The expected effect in the semiparametric model is computed using \eqref{eq_estimator_avg}. The effect on dispersion is computed using \eqref{eq_estimator_sd}. Standard errors in the semiparametric model are constructed via the delta method. Line ``p-value overid test'' reports the p-value for the specification test discussed in Online \Cref{app_asymptotic}
            \end{minipage}
            
\endgroup
\end{table}

    % latex table generated in R 4.3.1 by xtable 1.8-4 package
% Thu Oct 26 21:34:08 2023
\begin{table}[H]
\centering
\caption{Treatment effects on non-integrated rides: regular train/metro user} 
\label{table_user_unimodal}
\begingroup\footnotesize
\scalebox{0.55}{
\begin{tabular}{lllllllllllllllllll}
  &  & 2018-11-26 & 2018-12-10 & 2018-12-24 & 2019-01-07 & 2019-01-21 & 2019-02-04 & 2019-02-18 & 2019-03-04 & 2019-03-18 & 2019-04-01 & 2019-04-15 & 2019-04-29 & 2019-05-13 & 2019-05-27 & 2019-06-10 & 2019-06-24 & 2019-07-08 \\ 
  \hline \\ & &\multicolumn{ 17 }{c}{\textit{ Non-integrated rides }}  \\ \cmidrule(lr){3- 19 }\textit{Control group} &  &  &  &  &  &  &  &  &  &  &  &  &  &  &  &  &  &  \\ 
   & Mean & 2.2252 & 2.1450 & 1.2417 & 1.9587 & 2.1108 & 2.1144 & 2.2653 & 1.9045 & 2.0472 & 2.1356 & 1.9741 & 1.8762 & 1.7252 & 1.8361 & 1.7358 & 1.8137 & 1.5189 \\ 
   &  & (0.0185) & (0.0190) & (0.0125) & (0.0188) & (0.0191) & (0.0199) & (0.0203) & (0.0179) & (0.0198) & (0.0206) & (0.0191) & (0.0195) & (0.0183) & (0.0192) & (0.0176) & (0.0188) & (0.0163) \\ 
   &  &  &  &  &  &  &  &  &  &  &  &  &  &  &  &  &  &  \\ 
  \textit{20\% discount} &  &  &  &  &  &  &  &  &  &  &  &  &  &  &  &  &  &  \\ 
   & \textit{Nonparametric model} &  &  &  &  &  &  &  &  &  &  &  &  &  &  &  &  &  \\ 
   &  &  &  &  &  &  &  &  &  &  &  &  &  &  &  &  &  &  \\ 
   & Expected effect (levels) & 0.1331  & 0.0498  & 0.1636  & 0.1885  & -0.0446  & -0.1388  & -0.1481  & -0.1679  & -0.0999  & -0.2007  & -0.2063  & 0.0953  & 0.0502  & -0.0452  & 0.0564  & 0.0144  & 0.2389  \\ 
   &  & (0.2030)  & (0.1807)  & (0.1417)  & (0.1939)  & (0.2042)  & (0.2110)  & (0.2216)  & (0.1790)  & (0.2276)  & (0.2126)  & (0.2054)  & (0.2098)  & (0.2031)  & (0.1996)  & (0.1875)  & (0.2241)  & (0.2087)  \\ 
   & Effect on dispersion (relative) & -0.0163  & -0.1403  & 0.0442  & -0.0059  & -0.0916  & -0.1186  & -0.0758  & -0.2532 *** & -0.1430  & -0.1681  & -0.1989 ** & -0.0938  & -0.0984  & -0.1138  & 0.0441  & 0.0285  & 0.1512  \\ 
   &  & (0.0976)  & (0.0913)  & (0.1220)  & (0.1178)  & (0.0922)  & (0.1074)  & (0.0987)  & (0.0913)  & (0.0984)  & (0.1031)  & (0.0958)  & (0.1174)  & (0.1249)  & (0.1033)  & (0.1404)  & (0.1420)  & (0.1494)  \\ 
   &  &  &  &  &  &  &  &  &  &  &  &  &  &  &  &  &  &  \\ 
   & \textit{Semiparametric model (R=7)} &  &  &  &  &  &  &  &  &  &  &  &  &  &  &  &  &  \\ 
   &  &  &  &  &  &  &  &  &  &  &  &  &  &  &  &  &  &  \\ 
   & Expected effect (levels) & -0.0156  & -0.1647  & 0.1546  & 0.0028  & -0.0921  & -0.1834  & -0.2262  & -0.2420  & -0.1966  & -0.2786  & -0.3808 *** & -0.1612  & -0.1211  & -0.2150  & 0.0996  & -0.2290  & 0.0635  \\ 
   &  & (0.1676)  & (0.1632)  & (0.1112)  & (0.1698)  & (0.1647)  & (0.1715)  & (0.1926)  & (0.1560)  & (0.1686)  & (0.1779)  & (0.1384)  & (0.1911)  & (0.1665)  & (0.1576)  & (0.1585)  & (0.1958)  & (0.1599)  \\ 
   & Effect on dispersion (relative) & -0.0075  & -0.0895  & 0.1423  & 0.0018  & -0.0492  & -0.1001  & -0.1050  & -0.1396 * & -0.1022  & -0.1501 * & -0.2029 *** & -0.0931  & -0.0758  & -0.1273  & 0.0684  & -0.1220  & 0.0421  \\ 
   &  & (0.0805)  & (0.0834)  & (0.1095)  & (0.1033)  & (0.0855)  & (0.0880)  & (0.0834)  & (0.0808)  & (0.0818)  & (0.0871)  & (0.0633)  & (0.1034)  & (0.1000)  & (0.0854)  & (0.1120)  & (0.0967)  & (0.1081)  \\ 
   & p-value overid test & 0.3311  & 0.0054 *** & 0.1315  & 0.4471  & 0.5614  & 0.9722  & 0.5583  & 0.1532  & 0.6324  & 0.7526  & 0.3476  & 0.1504  & 0.0725 * & 0.5891  & 0.6963  & 0.3141  & 0.6933  \\ 
   &  &  &  &  &  &  &  &  &  &  &  &  &  &  &  &  &  &  \\ 
  \textit{50\% discount} &  &  &  &  &  &  &  &  &  &  &  &  &  &  &  &  &  &  \\ 
   & \textit{Nonparametric model} &  &  &  &  &  &  &  &  &  &  &  &  &  &  &  &  &  \\ 
   &  &  &  &  &  &  &  &  &  &  &  &  &  &  &  &  &  &  \\ 
   & Expected effect (levels) & -0.1632  & 0.0792  & 0.0173  & -0.0767  & -0.2081  & -0.2667  & -0.3664 * & -0.3688 ** & -0.2787  & -0.3140  & -0.3436 * & -0.1601  & -0.0775  & -0.1232  & -0.0201  & -0.2179  & -0.1791  \\ 
   &  & (0.1971)  & (0.1901)  & (0.1455)  & (0.1870)  & (0.2006)  & (0.2050)  & (0.2152)  & (0.1840)  & (0.2203)  & (0.2166)  & (0.2022)  & (0.2096)  & (0.1972)  & (0.2027)  & (0.1786)  & (0.2157)  & (0.1911)  \\ 
   & Effect on dispersion (relative) & -0.1076  & -0.0669  & 0.0629  & -0.1097  & -0.1568 * & -0.2050 ** & -0.1648 ** & -0.2213 ** & -0.2367 *** & -0.1573  & -0.2571 *** & -0.1251  & -0.1868 * & -0.1117  & -0.0862  & -0.0806  & -0.0608  \\ 
   &  & (0.1019)  & (0.1039)  & (0.1456)  & (0.1097)  & (0.0868)  & (0.0964)  & (0.0840)  & (0.0973)  & (0.0912)  & (0.1079)  & (0.0919)  & (0.1152)  & (0.0981)  & (0.1165)  & (0.1056)  & (0.1328)  & (0.1405)  \\ 
   &  &  &  &  &  &  &  &  &  &  &  &  &  &  &  &  &  &  \\ 
   & \textit{Semiparametric model (R=4)} &  &  &  &  &  &  &  &  &  &  &  &  &  &  &  &  &  \\ 
   &  &  &  &  &  &  &  &  &  &  &  &  &  &  &  &  &  &  \\ 
   & Expected effect (levels) & -0.1374  & -0.0070  & 0.0081  & -0.0833  & -0.2880  & -0.2891  & -0.3361 * & -0.2989 * & -0.3907 * & -0.2658  & -0.3223 * & -0.1262  & -0.1882  & -0.1128  & -0.0752  & -0.2145  & -0.2230  \\ 
   &  & (0.1867)  & (0.1804)  & (0.1341)  & (0.1688)  & (0.1840)  & (0.1924)  & (0.1947)  & (0.1679)  & (0.2148)  & (0.2026)  & (0.1923)  & (0.1933)  & (0.1860)  & (0.1900)  & (0.1659)  & (0.2018)  & (0.1824)  \\ 
   & Effect on dispersion (relative) & -0.0672  & -0.0071  & 0.0085  & -0.0543  & -0.1506 * & -0.1593 * & -0.1602 * & -0.1809 ** & -0.2007 ** & -0.1481  & -0.1833 * & -0.0808  & -0.1217  & -0.0710  & -0.0487  & -0.1239  & -0.1543  \\ 
   &  & (0.0894)  & (0.1000)  & (0.1232)  & (0.1049)  & (0.0880)  & (0.0949)  & (0.0840)  & (0.0904)  & (0.0961)  & (0.1037)  & (0.0956)  & (0.1167)  & (0.1090)  & (0.1142)  & (0.1065)  & (0.1082)  & (0.1157)  \\ 
   & p-value overid test & 0.4841  & 0.3120  & 0.9709  & 0.6105  & 0.6587  & 0.7319  & 0.9262  & 0.3001  & 0.2214  & 0.7459  & 0.0586 * & 0.2249  & 0.1256  & 0.2505  & 0.4432  & 0.6968  & 0.5752  \\ 
   \hline
\end{tabular}
}
\endgroup
\begin{minipage}{0.9\linewidth}
 \vspace{1em}
%       \begin{minipage}{\textwidth} -- what should go in this width argument?
            \tiny \textit{Notes:} $***$: $p < 0.01$; $**$: $p < 0.05$;  $*$: $p < 0.10$. Standard error in parentheses. Sample sizes for each experimental arm are:  387 (control), 406 (20\% discount) and 380 (50\% discount). Estimates for expected effects under ``Nonparametric model'' correspond to difference-in-means estimators, which are efficient when no restrictions on potential outcomes are imposed \citep{Newey1990}. Estimates for effects on dispersion under ``Nonparametric model'' are based on the ratio of estimated standard deviations minus one, with standard errors computed via the delta method. Estimates under ``Semiparametric model'' are constructed using model \eqref{eq_model_n} and estimator \eqref{eq_estimator_text}, with optimal weighting scheme estimated according to the weighted bootstrap in Online Appendix \ref{app_asymptotic}, no trimming, and $R$ selected using pre-treatment data (Online Appendix \ref{app_selection}). The expected effect in the semiparametric model is computed using \eqref{eq_estimator_avg}. The effect on dispersion is computed using \eqref{eq_estimator_sd}. Standard errors in the semiparametric model are constructed via the delta method. Line ``p-value overid test'' reports the p-value for the specification test discussed in Online \Cref{app_asymptotic}
            \end{minipage}
\end{table}

\end{landscape}

% latex table generated in R 4.3.1 by xtable 1.8-4 package
% Mon Oct 30 19:36:30 2023
\begin{table}[H]
\centering
\caption{Contemporaneous effects on integrated rides} 
\label{table_results_calibration}
\scalebox{0.8}{
\begin{tabular}{lllllll}
  \hline
  & \multicolumn{2}{c}{All} &\multicolumn{2}{c}{Regular train/metro user} & \multicolumn{2}{c}{Not regular train/metro user} \\  \cmidrule(lr){2-3}
\cmidrule(lr){4-5} \cmidrule(lr){6-7} & sample avg effect & \% learning & sample avg effect & \% learning &  sample avg effect & \% learning \\ \\ \textit{20\% discount}& & & & & & \\    \hline
Estimate & 0.0457 & -5.2532 & 0.0919 & 20.0476 & 0.0044 & -477.3836 \\ 
   & (0.0291) & (3.3503) & (0.0559) & (12.1944) & (0.0234) & (2543.4938) \\ 
   Sample size &  \multicolumn{2}{c}{ 1538 }&\multicolumn{2}{c}{ 726 }&\multicolumn{2}{c}{ 812 }  \\ \hline  \\  \textit{50\% discount}& & & & & & \\ \hline  Estimate & 0.0871*** & 42.7929*** & 0.1193* & 38.9193* & 0.0586*** & 49.7826*** \\ 
   & (0.0310) & (15.2127) & (0.0619) & (20.1825) & (0.0202) & (17.1253) \\ 
   N &  \multicolumn{2}{c}{ 1490 }&\multicolumn{2}{c}{ 700 }&\multicolumn{2}{c}{ 790 }  \\  \hline  \\     Avg (control) &  \multicolumn{2}{c}{ 0.1419 }&\multicolumn{2}{c}{ 0.2228 }&\multicolumn{2}{c}{ 0.067 } \\ \hline
\end{tabular}
}

\begin{minipage}{1.0\linewidth}
 \vspace{1em}
%       \begin{minipage}{\textwidth} -- what should go in this width argument?
            \scriptsize \textit{Notes:} $***$: $p < 0.01$; $**$: $p < 0.05$;  $*$: $p < 0.10$. Standard error in parentheses.The sample average treatment effect for regime $d \in \{20\%,50\%\}$ is defined as $\mathbf{T}_d = \frac{1}{|\mathcal{N}_d \cup \mathcal{N}_0|}\sum_{i \in \mathcal{N}_d \cup \mathcal{N}_0}(b_{\text{November, 26th}}(d;U_i) - b_{\text{November, 26th}}(d;U_i))$, where $\mathcal{N}_s$ is the set of individuals assigned to experimental group $s$ for which income and travel time data is available at the baseline. Subsample average effects are similarly defined. These parameters are estimated using the semiparametric model. Standard errors are computed using the Delta Method \citep[see][]{Athey2023}. The share attributable to learning relies on the calibration discussed in \Cref{sec_decomposition}. The main calibrated parameters are: (i) prior mean of $\mathbb{E}[\log(A)|\mathcal{H}_{\text{November, 26th}}] = 0.487$; (ii) prior variance of $\mathbb{V}[\log(A)|\mathcal{H}_{\text{November, 26th}}] = 0.785$; (iii) $\gamma = 4.055$; (iv) $\phi = 0.514$; (v) (full sample average) $\lambda$ $=9.158563$. Standard errors for the share attributable to learning are computed using the Delta Method, abstracting from uncertainty in the calibration. 
            \end{minipage}
\end{table}

\bibliography{bibliography}
\bibliographystyle{apalike}

\addresseshere

\pagebreak
\setcounter{page}{1}

\section*{Online Appendix to ``The learning effects of subsidies to bundled goods: a semiparametric approach''}
\label{supplementary}
\appendix
\section{Proof of \Cref{prop_learning}}
\label{proof_prop}
\begin{proof}
We prove each item separately:

\paragraph{1} Fix signal precisions  $h'', h' \in \operatorname{Im}(\bar{h})$, $h' < h''$. Denote by $Z_t(h)$ the signal under precision $h$. By the stability assumption, $\xi_t \coloneqq Z_t(h') - Z_t(h'')$ is independent of $Z_t(h'')$, conditionally on $\sigma(\mathcal{H}_t,A)$. Moreover, since $\xi_t$ is distributed as $G(\frac{1}{\rho} \cdot)$ for $\rho > 0$, conditionally on $\sigma(\mathcal{H}_t,A)$, it follows that $\xi_t$ is independent of $\sigma(\mathcal{H}_t,A)$. Consequently, defining $\tilde{\mathcal{H}}_{t+1}(h'') = \sigma(\mathcal{H}_t, Z_t(h''), \xi_t)$, we have that $\mathbb{E}[g(A)|\mathcal{H}_{t+1}(h'')]$ is a version of $\mathbb{E}[g(A)|\tilde{\mathcal{H}}_{t+1}(h'')]$ for any $g$ s.t. $\mathbb{E}[|g(A)|]<\infty$.\footnote{\textit{Proof:} Let $X$ be a version of $\mathbb{E}[g(A)|\mathcal{H}_{t+1}(h'')]$. Observe that $X$ is $\tilde{\mathcal{H}}_{t+1}(h'')$-measurable. Fix $B \in \mathcal{H}_t$, $C \in \sigma({Z}_t(h''))$ and $D \in \sigma(\mathcal{\xi}_t)$. Note that:
$$\mathbb{E}[g(A)\mathbf{1}_{B\cap C \cap D}] = \mathbb{E}[g(A)\mathbf{1}_{B} \mathbb{E}[\mathbf{1}_{C\cap D}|\sigma(A,\mathcal{H}_t)]] = \mathbb{E}[g(A)\mathbf{1}_{B\cap C}]\mathbb{P}[D]=\mathbb{E}[X\mathbf{1}_{B\cap C}]\mathbb{P}[D] = \mathbb{E}[X\mathbf{1}_{B\cap C\cap D}] \, ,$$
where the first equality follows from iterated expectations, the second equality follows from the independence conditions, the third equality follows from the definition of $\mathbb{E}[g(A)|\mathcal{H}_{t+1}(h'')]$, and the final equality follows from $\mathbb{E}[\mathbf{1}_{D}|\mathcal{H}_{t+1}(h'')] = \mathbb{E}[\mathbb{E}[\mathbf{1}_{D}|\sigma({\mathcal{H}}_{t+1}(h''),A)]|\mathcal{H}_{t+1}(h'')] = \mathbb{P}[D]$, which is true due to the independence conditions. The conclusion then follows from Theorem 34.1 in \cite{Billingsley2012}.} Consequently, $\mathbb{E}[\mathbb{E}[g(A)|\mathcal{H}_{t+1}(h'')]|\mathcal{H}_{t+1}(h')] = \mathbb{E}[g(A)|\mathcal{H}_{t+1}(h')]$, and iterated expectations leads to:

\begin{equation*}
    \begin{aligned}    
    \mathbb{E}[V_{t+1}(a_{t+1}, \mathcal{H}_{t+1}(h'))|\mathcal{H}_t]= \\
    \mathbb{E}\Big[ \max_{o,b,n,c\geq 0, a' \geq \underline{a}_{t+2}}  \Big\{ \mathbb{E}\Big[  u_{t+1}(c,\tau|\mathcal{H}_{t+1}(h'')) + \beta \mathbb{E}[V_{t+2}(a',\mathcal{H}_{t+2}(h', \bar{h}(b,n))|\mathcal{H}_{t+1}(h'')]\Big|\mathcal{H}_{t+1}(h')\Big] \Big\} \Big|\mathcal{H}_t\Big] \leq \\ \mathbb{E}\Big[ \max_{o,b,n,c\geq 0, a' \geq \underline{a}_{t+2}} u_{t+1}(c,\tau|\mathcal{H}_{t+1}(h'')) + \beta \mathbb{E}[V_{t+2}(a',\mathcal{H}_{t+2}(h', \bar{h}(b,n)))|\mathcal{H}_{t+1}(h'')] \Big|\mathcal{H}_t\Big]\, ,
    \end{aligned}
\end{equation*}
where $\mathcal{H}_{t+2}(h', \bar{h}(b,n)) = \sigma(\mathcal{H}_t, Z_t(h'), Z_{t+1}(\bar{h}(b,n)))$. By a similar reasoning, we are able to show that, for any $\tilde{a}_{t+2}\geq \underline{a}_{t+2}$, $h_{t+2}\in \operatorname{Im}(\bar{h})$:

\begin{equation*}
\begin{aligned}
  \mathbb{E}[V_{t+2}(\tilde{a}_{t+2},\mathcal{H}_{t+2}(h', h_{t+2}))|\mathcal{H}_{t+1}(h')] \leq \\ \mathbb{E}\Big[ \max_{o,b,n,c\geq 0, a' \geq \underline{a}_{t+3}} u_{t+2}(c,\tau|\mathcal{H}_{t+2}(h'', h_{t+2})) + \beta \mathbb{E}[V_{t+2}(a',\mathcal{H}_{t+3}(h',  h_{t+2}, \bar{h}(b,n)))|\mathcal{H}_{t+2}(h'', h_{t+2})] \Big|\mathcal{H}_{t+1}(h'))\Big] \, ,
\end{aligned}
\end{equation*}
from which we are able to conclude that:
\begin{equation*}
    \begin{aligned}       
    \mathbb{E}[V_{t+1}(a_{t+1}, \mathcal{H}_{t+1}(h'))|\mathcal{H}_t]\leq \\ 
    \mathbb{E}\Big[ \max_{o,b,n,c\geq 0, a' \geq \underline{a}_{t+2}} \Big\{ u_{t+1}(c,\tau|\mathcal{H}_{t+1}(h'')) + \beta \mathbb{E}\Big[\max_{o',b',n',c'\geq 0, a'' \geq \underline{a}_{t+2}} \Big\{ u_{t+2}(c',\tau'|\mathcal{H}_{t+2}(h'', \bar{h}(b,n)))+ \\ \beta \mathbb{E}[V_{t+2}(a'',\mathcal{H}_{t+3}(h', \bar{h}(b,n), \bar{h}(b',n')))|\mathcal{H}_{t+2}(h'', \bar{h}(b,n))] \Big\} \Big|\mathcal{H}_{t+1}(h'')\Big]\Big\}\Big|\mathcal{H}_t\Big]\, ,
    \end{aligned}
    \end{equation*}

Proceeding inductively,
and then using that, by the assumptions of the problem, $$\left|\beta^s V_{t+s}(\tilde{a}_{t+s}, \mathcal{H}_{t+s}(h', h_{t+2}, \ldots h_{t+s}) )\right| \leq \frac{\beta^s}{1-\beta}\max\left\{\bar{w}, \frac{\lambda}{2}\bar{\tau}^2 \right\} + \beta^{s-1}|\tilde{a}_{t+1}| \, ,$$ 
for any optimal sequences of precisions
$(h_{t+s})_{s\geq1}$ and wealth  $(\tilde{a}_{t+s})_{s\geq 1}$ starting from $h_{t+1}=h'$ and $\tilde{a}_{t+1}=a_{t+1}$; where $\bar{w} = \sup_{n}w_n$ and $\bar{\tau} = \sup_{n} \tau_n^*$; and, by the transversality condition, $\lim_{s \to \infty} \beta^s|\tilde{a}_{t+1}|_{t+1} =0$, we arrive at the desired conclusion.

\paragraph{2} Let $\mathcal{F}_\infty = \sigma(\mathcal{H}_1, Z_1, Z_2, \ldots )$. For each $n \in \mathbb{N}$, define $\bar{Z}_n = \frac{1}{\sum_{j=1}^n{h}_j}\sum_{j=1}^n{h_j} Z_j = \psi(A) + \frac{1}{\sum_{j=1}^n{h}_j}\sum_{j=1}^n V_j$. Observe that, for each $n \in \mathbb{N}$, $\bar{Z}_n$ is $\mathcal{F}_\infty$-measurable; thus, $\limsup_{n \to \infty} \bar{Z}_n$ is also $\mathcal{F}_\infty$-measurable. But, by the strong law of large numbers and the learning assumption, $\limsup_{n \to \infty} \bar{Z}_n = \psi(A)$, which implies that $\psi(A)$, and, by measurability of $\psi^{-1}$, $A$, are $\mathcal{F}_\infty$-measurable. The desired conclusion then follows from Theorem 4.6.8 in \cite{Durrett2019}.

\paragraph{3} At an interior optimum, the solution is characterised by the linear system:

\begin{equation}
\label{eq_foc}
    \begin{bmatrix}
        1 & \gamma \mathbb{E}[A^\phi|\mathcal{H}_t] &  \mathbb{E}[A^|\mathcal{H}_t] \\
        \gamma \mathbb{E}[A^\phi|\mathcal{H}_t] & \gamma^2 \mathbb{E}[A^{2\phi}|\mathcal{H}_t] & \gamma \mathbb{E}[A^{1+\phi}|\mathcal{H}_t] \\ 
            \mathbb{E}[A|\mathcal{H}_t] & \gamma \mathbb{E}[A^{1 + \phi}|\mathcal{H}_t] & \mathbb{E}[A^{2}|\mathcal{H}_t]
    \end{bmatrix} \begin{bmatrix}
        r_t \\
        b_t \\
        n_t
    \end{bmatrix} =\begin{bmatrix}
    \tau_t^* - \frac{1}{\lambda}p_{o,t} \\
    \tau_t^*\gamma \mathbb{E}[A^\phi|\mathcal{H}_t]  - \frac{1}{\lambda}p_{b,t} + \frac{\beta}{\lambda} \frac{\partial \mathbb{E}[V_{t+1}|\mathcal{H}_t]}{\partial \bar{h}} \frac{\partial \bar{h}_t}{\partial b_t} \\
    \tau_t^* \mathbb{E}[A|\mathcal{H}_t]  - \frac{1}{\lambda}p_{n,t} + \frac{\beta}{\lambda} \frac{\partial \mathbb{E}[V_{t+1}|\mathcal{H}_t]}{\partial \bar{h}_t} \frac{\partial \bar{h}_t}{\partial n_t} \, ,
    \end{bmatrix}
\end{equation}
where, by assumption, the left-hand-side matrix is invertible. The conclusion then follows.
\paragraph{4} The conclusion follows from noting that, first, at an interior optimum, \eqref{eq_foc} is satisfied at each period. Moreover, by the envelope condition, the marginal value of information satisfies the recursion, for  given $s \in \mathbb{N}$ and every $k \geq s$:

\begin{equation}
   \mathbb{E}\left[ \frac{\partial V_{s+k}}{\partial \bar{h}_s} \Big|\mathcal{H}_s\right]= \mathbb{E}\left[\frac{\partial u_{s+k}(c_{s+k},\tau_{s+k}|\mathcal{H}_{s+k})}{\partial \bar{h}_s}\Big|\mathcal{H}_s\right] + \beta \mathbb{E}\left[\frac{\partial V_{s+k+1}}{\partial \bar{h}_s}\Big|\mathcal{H}_s\right] \, ,
\end{equation}
with terminal condition $\lim_{k \to \infty}\beta^k \mathbb{E}\left[\frac{\partial V_{s+k}}{\partial \bar{h}_s}\Big|\mathcal{H}_s\right] = 0 $. Consequently, if there is no change in the information set $\mathcal{H}_{t+1}$, the conditions for an optimum remain unchanged.

\end{proof}

\section{Details on semiparametric L-moment estimator}
\label{app_metrics}
In this Appendix, we present details on the generalised L-moment estimator discussed in the main text. In what follows, we consider the semiparametric model for scalar potential outcomes $Y(0)$ and $Y(1)$:

\begin{equation}
    \begin{aligned}
Y(1) = G(Y(0); \theta_0), \quad \theta_0 \in \Theta \subseteq \mathbb{R}^p \, , 
    \end{aligned}
\end{equation}
where $G$ is a known function up to $\theta_0$ with the property that, for each $\theta \in \Theta$, $u \mapsto G(u;\theta)$ is increasing in $u$. 

Suppose we have access to two independent random samples from, respectively, the distributions of $Y(0)$ and $Y(1)$.\footnote{This is the case in simple randomised experiments as the one considered in the main text.} To estimate $\theta_0$, we consider a semiparametric version of the generalised method of L-moments (GMLM) estimator of \cite{alvarez2023inference}, henceforth ACM. Specifically, we propose to estimate $\theta$ by:

\begin{equation}
\label{eq_estimator}
    \begin{aligned}
       \hat{\theta} \in \operatorname{argmin}_{\theta \in \Theta} \left\lVert \int_{\underline{p}}^{\overline{p}} (\hat{Q}_{1}(u) -  G(\hat{Q}_{0}(u);\theta)) \boldsymbol{P}_R(u) du \right\rVert_{W_R}^2 \, ,
    \end{aligned}
\end{equation}
where $\hat{Q}_d$ is the empirical quantile function of the random sample from $Y(d)$, $d \in \{0,1\}$; $\boldsymbol{P}_R(u) = (P_1(u),\ldots, P_{R}(u))'$ is a vector of $R$ basis functions, with the property that $\{P_r\}_{r \in \mathbb{N}}$ form an orthonormal basis on $L^2[0,1]$;\footnote{The Legendre polynomials in the definition of L-moments given in \Cref{sec_estimation} satisfy this requirement up to rescaling.} $0 \leq \underline{p} \leq \overline{p}\leq 1$ are trimming constants; and, for a vector $x \in \mathbb{R}^R$, $\lVert x \rVert_{W_R} = \sqrt{x'W_Rx}$, where $W_R$ is a (possibly estimated) $R\times R$ weighting matrix.

In the next sections, we discuss the asymptotic properties of our proposed estimator, provide conditions under which it is efficient, propose a method to tune $R$ and, possibly, $\underline{p}$ and $\overline{p}$, and assess the properties of our proposed estimator in a Monte Carlo exercise.

\subsection{Asymptotic normality, optimal weighting scheme and overidentification test}
\label{app_asymptotic}

Let $Q_d$ denote the quantile function of the distribution of $Y(d)$, $d \in \{0,1\}$. Denote by $N_d$ the sample size of the random sample from $Y(d)$, and $N = N_0 + N_1$ the full sample size. Under standard identifiability and differentiability conditions (see Propositions 1 and 2 of ACM), we are able to show that, if $N_1, N_0 \to \infty$ with $\lim_{n \to \infty} N_0/N = p_0 \in (0,1)$, and, for a sequence $R_{N} \to \infty$ of basis functions, the following hold:

\begin{enumerate}
    \item (asymptotic tightness) for $d \in \{0,1\}$, $\int_{\underline{p}}^{\overline{p}}
(\sqrt{n}(\hat{Q}_d(u) - Q_d(u))^2 du = O_{\mathbb{P}}(1)$;\footnote{This condition is implied by weak convergence of the empirical quantile process on $L_{q}[\underline{p},\overline{p}]$, where $p \in [2,\infty]$.}
\item (consistent estimation of weighting matrix) there exists a sequence of positive semidefinite $R_N \times R_N$ matrices $\Omega_{R_N}$ such that $\lVert W_{R_N} - \Omega_{R_N} \rVert_{2} = o_{\mathbb{P}}(1)$, where $\lVert \cdot \rVert_2$ denotes the spectral norm;
\item (rank condition) definining the Jacobian $J_{R_N}(\theta) \coloneqq -\int_{\underline{p}}^{\overline{p}} \nabla_{\theta'} G(u;\theta) \boldsymbol{P}_{R_N}(u) du$, the eigenvalues of $ M_{R_N} \coloneqq J_{R_N}(\theta_0)'\Omega_{R_N} J{R_N}(\theta_0)$ are uniformly bounded away from zero;\footnote{Similarly to the finite moment case \citep{Rothenberg1971}, this condition is related to identification. See Appendix B of ACM.}
\end{enumerate}
then the estimator admits the following asymptotic linear representation:\footnote{See the proofs of Propositions 1 and 2 in ACM for details. The construction follows the standard argument for asymptotic linearity of generalised method of moments estimators with a finite number moments \citep{Newey1994}, while repeatedly relying on the following corollary of Bessel's inequality \citep{Kreyszig1989} to account for the increasing number of L-moments: for any $\xi\in L_2[\underline{p},\overline{p}]$, $\lVert\int_{\underline{p}}^{\overline{p}} \xi(u) \boldsymbol{P}_{R_n}(u) du \rVert^2_2 \leq \int_{\underline{p}}^{\overline{p}} \xi(u)^2 du$.}

\begin{equation}
\footnotesize
    \begin{aligned}
        \sqrt{N}(\hat{\theta} - \theta_0) = \\ - M_{R_N}^{-1}J_{R_N}(\theta_0)'\Omega_{R_N}\int_{\underline{p}}^{\overline{p}}\left(\sqrt{\frac{N_1}{p_1}} (\hat{Q}_1(u) - Q_1(u)) - \partial_{Y}G(Q_{1}(u);\theta_0)\sqrt{\frac{N_0}{p_0}} (\hat{Q}_0(u) - Q_0(u))\right)\boldsymbol{P}_{R_N}(u) du  + o_{\mathbb{P}}(1)\, ,
    \end{aligned}
\end{equation}

This representation may be used as a basis for inference. Specifically, if, for each $d \in \{0,1\}$, $Y(d)$ admits Lebesgue density $f_d$ that, when restricted to $[Q_{d}(\underline{p}), Q_{d}(\overline{p})]$, satisfies the Assumptions of Theorem 5 of \cite{Csorgo1978}, then, on the same probability space of $\hat{Q}_0$ and $\hat{Q}_1$, one can define two independent sequences of Brownian bridges $\{B_{n}^0\}_{n \in \mathbb{N}}$, $\{B_{n}^1\}_{n \in \mathbb{N}}$  such that:

$$\sup_{u \in [\underline{p},\overline{p}]} \left|\sqrt{N}_d(\hat{Q}_d(u) - Q_d(u)) - f_d(Q_d(u))^{-1} B_{N_d}^d(u) \right| = o_{\mathbb{P}}(1) \, , \quad  d \in \{0,1\} \, .$$

Consequently, if $\int_{\underline{p}}^{\overline{p}} \partial_Y G(Q_1(u);\theta_0)^2 du < \infty$, an application of Bessel's inequality shows that:

\begin{equation}
\footnotesize
\label{asy_strong}
    \begin{aligned}
        \sqrt{N}(\hat{\theta} - \theta_0) = \\ - M_{R_N}^{-1}J_{R_N}(\theta_0)'\Omega_{R_N}\int_{\underline{p}}^{\overline{p}}\left(\frac{B_{N_1}^1(u) }{\sqrt{p_1}f_1(Q_1(u))} - \partial_{Y}G(Q_{1}(u);\theta_0)\frac{B_{N_0}^0(u) }{\sqrt{p_0}f_0(Q_0(u))}\right)\boldsymbol{P}_{R_N}(u) du  + o_{\mathbb{P}}(1)\, .
    \end{aligned}
\end{equation}

Now, let $V_N$ be the variance of the leading term of the right-hand side in \eqref{asy_strong}. Since the leading term is Gaussian, Slutsky's theorem shows that $V_{N}^{-1/2}  \sqrt{N}(\hat{\theta} - \theta_0) \overset{d}{\to} N(0, \mathbb{I}_p)$, thus establishing asymptotic normality of the generalised L-moment estimator.

\begin{remark}[On the optimal weigthing scheme]
From \eqref{asy_strong}, we have that the weighting matrix that minimises the variance of the approximation on the right-hand side is:

\begin{equation}
\label{eq_optimal}
        \Omega_{R_N}^* = \mathbb{V}\left[\int_{\underline{p}}^{\overline{p}}\left(\frac{B_{N_1}^1(u) }{\sqrt{p_1}f_1(Q_1(u))} - \partial_{Y}G(Q_{1}(u);\theta_0)\frac{B_{N_0}^0(u) }{\sqrt{p_0}f_0(Q_0(u))}\right)\boldsymbol{P}_{R_N}(u) du \right]^{-} \, ,
\end{equation}
where $A^-$ denotes the generalised inverse of a matrix $A$. Following the generalised method of moments literature \citep{Wooldridge2010}, this matrix may be estimated in two-steps, by first estimating  $\theta_0$ using \eqref{eq_estimator} and a known weighting scheme and then plugging it in the formula given by \eqref{eq_optimal}.\footnote{In some settings, the derivative  $\partial_{Y}G(Q_{1}(u);\theta_0)$ either does not depend on $\theta_0$, in which case preliminary estimation of $\theta_0$ via the GMLM is unnecessary; or depends on a subset of $\theta_0$ which can be estimated nonparametrically, in which case the first step may also proceed without resorting to the GMLM. As an example of the former, we have the location model $Y(1) = \theta_{0,0} + Y(1)$, which is analyzed in the Monte Carlo exercise (\Cref{app_montecarlo}). As an example of the latter, we have the location-scale model of the main text, $Y(1) = \theta_{0,0} + \theta_{0,1}Y(1)$, where the partial derivative depends solely on $\theta_{0,1}$, which can be estimated nonparametrically in the first-step by $\sqrt{\hat{\mathbb{V}[Y(1)]}/\hat{\mathbb{V}[Y(0)]}}$.} The density $f_0$ may be estimated nonparametrically, whereas $f_1$ may be estimated non- or semiparametrically. However, instead of directly estimating $f_d$, we suggest an approach based on the weighted bootstrap, which takes a convenient form and nests the Bayesian bootstrap of \cite{Rubin1981} as a specififc case. Specifically, let $\tilde{F}_d(x) = \sum_{i=1}^{N_d} w_i \mathbf{1}\{Y_{i,d}\leq x \} $ be a randomly reweighted version of the empirical cdf of the sample from $Y(d)$, where the weights $w_i$ are generated as $w_i = z_i/(\sum_{j=1}^{N_1} z_i)$, and $z_i$ is drawn iid from a distribution function with zero mean, unit variance and finite moment-generating function, independently from the data. Let  $\tilde{Q}_d(u)$ denote the empirical quantile function obtained from this reweighted cdf. Under \citeauthor{Csorgo1978}-type restrictions, \cite{AlvarezAndrade2013} show that there exists a sequence of Brownian bridges $\{\tilde{B}_n^d\}_{n \in \mathbb{N}}$ such that:

$$\sup_{u \in [\underline{p}. \overline{p}]}|\sqrt{N_d}(\tilde{Q}_d(u) - \hat{Q}_d(u)) - f_d(Q_d(u))^{-1}\tilde{B}^{d}_{N_d}(u)| \overset{a.s.}{=} O\left(\frac{\log(n)}{\sqrt{n}}\right) \, , $$
where, upon inspection of the proof of Theorem 4 of \cite{AlvarezAndrade2013}, the Brownian bridges may be taken as independent from the random samples from $Y(0)$ and $Y(1)$. Moreover, denoting by $\mathbb{P}^*$ the conditional on the sample law, we have, by Markov inequality:

$$\mathbb{P}^*\left[\sup_{u \in [\underline{p}. \overline{p}]}|\sqrt{N_d}(\tilde{Q}_d(u) - \hat{Q}_d(u)) - f_d(Q_d(u))^{-1}\tilde{B}^{d}_{N_d}(u)|  > \epsilon \right]= o_\mathbb{P}(1) \, , \quad \forall \epsilon > 0 \, .$$

The above approximation suggests using the weighted bootstrap to estimate the moments required by \eqref{eq_optimal}. The consistency of the bootstrap moment estimates then follows from the approximation above, coupled with uniform integrability conditions \citep{Kato2011}.

\end{remark}

\begin{remark}{(A test of overidentifying restrictions)} Following the GMM literature, whenever $R_{N} > p $, we may consider a $J$-test for overidentifying restrictions \citep{Wooldridge2010}:

\begin{equation}
    \label{eq_jtest}
\hat{J}_N = N  \left\lVert \int_{\underline{p}}^{\overline{p}} (\hat{Q}_{1}(u) -  G(\hat{Q}_{0}(u);\hat{\theta})) \boldsymbol{P}_{R_N}(u) du \right\rVert_{W_{R_N}}^2 \, . 
\end{equation}

Using the Gaussian approximations discussed above, we can show that, under the null:

$$H_0: \exists \theta \in \Theta, Y(1) = G(Y(0);\theta) \, ,$$
and when a consistent estimator of the optimal weighting scheme is used,
there exists a sequence $\{x_n\}_{n \in \mathbb{N}}$ of random variables, where $x_n \sim \chi^2(R_n-p)$, such that:

\begin{equation}
	\label{eq_distributional}
	\hat{J}_N =  x_n + o_{\mathbb{P}}(1) \, .
\end{equation}

The above result suggests using the critical values of a $\chi^2(R_n - p)$ to conduct a test on the overidentifying restrictions. Indeed, by relying on an inequality due to \citep{Gotze2019}, ACM show that \eqref{eq_distributional} implies that the quantiles of the chi-squared distribution may be used as valid critical values for the test. See Supplemental Appendix D of ACM for details.
\end{remark}
\subsection{Semiparametric efficiency}

In this section, we show that, when no trimming is adopted, i.e. $0=\underline{p}\leq \overline{p}=1$, the optimally weighted L-moment estimator achieves the semiparametric efficiency bound of the model \eqref{eq_model}. To see this, define the alternative (unfeasible) estimator:

\begin{equation*}
\begin{aligned}
        \check{\theta} \in \operatorname{argmin}_{\theta \in \Theta} \sum_{i \in \mathcal{S}_N} \sum_{j \in \mathcal{S}_N}\left(\frac{1}{\sqrt{p}_1}(\hat{Q}_{1}(i)-Q_{1}(\theta)) + \frac{1}{\sqrt{p}_0}\partial_q     G\cdot (\hat{Q}_{0}(i)-Q_{0}(i))\right) \times \kappa_{i,j} \times \\ \left(\frac{1}{\sqrt{p}_1}(\hat{Q}_{1}(j)-Q_{1}(\theta)) + \frac{1}{\sqrt{p}_0}\partial_q     G\cdot (\hat{Q}_{0}(j)-Q_{0}(j))\right)  \, ,
\end{aligned}
\end{equation*}
for a grid of $S_N$ points $\mathcal{S}_N =\{s_1,s_2,\ldots, s_{S_N}\} \subseteq (0,1)$ and weights $\kappa_{i,j}$, $i,j \in \mathcal{S}_N$. Here, $Q_{1}(\theta) = G(Q_{0}(u),\theta)$ and $\partial_q G = \partial_Y G(Q_{0}(u), \theta_0)$. Under standard conditions, and as $N,S_N\to \infty$, the estimator has asymptotic linear representation as follows:

\begin{equation*}
\begin{aligned}
   \sqrt{N}(\check{\theta}-\theta_0) = \, \, &  -(\partial_{\theta'} G_{S_N} '\boldsymbol{\kappa}_{S_N} \partial_{\theta'} G_{S_N})^{-1} \partial_{\theta'} G_{S_N}' \boldsymbol{\kappa}_{G_N}\Big[ \frac{1}{\sqrt{p}_1}\boldsymbol{f_{1}}^{-1} * \sqrt{N_1}F_{Y(1),S_N} + \\ & \frac{\boldsymbol{\partial_q G}}{\sqrt{p}_0} * \boldsymbol{f_{0}}^{-1} * \sqrt{N_0}F_{Y(0),S_N}\Big] + o_{\mathbb{P}}(1) \, ,
\end{aligned}
\end{equation*}
where $*$ denotes entry-by-entry multiplication,  ${\boldsymbol{f}_{d}}^{-1} = \begin{pmatrix}
\frac{1}{f_{d}(Q_{Y(d)}(s_1)} & \frac{1}{f_{d}(Q_{Y(d)}(s_2))} & \ldots & \frac{1}{f_{d}(Q_{Y(d)}(s_{S_N}))}  
\end{pmatrix}'$ and $\boldsymbol{\partial_q G}$ is similarly defined. First, we observe that:

\begin{equation*}
   \frac{1}{f_{1}(Q_{1}(v))} = Q'_{Y(1)}(v) = \partial_q G(Q_{0}(v);\theta_0) \cdot \frac{1}{f_{0}(Q_{0}(v))} \, ,
\end{equation*}
which implies, by taking $\mathcal{S}_N = \Big\{\frac{1}{S_N+1}, \frac{2}{S_N+1}, \ldots, \frac{S_N}{S_N+1}\Big\}$ and applying Lemma C.1. in \cite{Firpo2021}, that we are able to show that the variance of the estimator under optimal weights is:

\begin{equation*}
    \mathbb{V}^* =  ((\partial G_{S_N} * ( \mathbf{1}_{d}' \otimes \boldsymbol{f_{1}}))' \Sigma_{S_N}^{-1} (\partial G_{S_N} * ( \mathbf{1}_{d}' \otimes \boldsymbol{f_{1}} )))^{-1} \, ,
\end{equation*}
where
\begin{equation*}
    (\Sigma_{S_N}^{-1})_{s_i,s_j} = p_0p_1 [\mathbbm{1}_{\{s_i = s_j\}} 2(S_N+1) - (\mathbbm{1}_{\{s_{i} = s_{j+1}\}}  + \mathbbm{1}_{\{s_{i} = s_{j-1}\}}) (S_N+1)] \, .
\end{equation*}

Proceeding similarly as in Appendix G of ACM, we obtain that:

\begin{equation*}
     \lim_{N \to \infty}\left(\mathbbm{V}^{*-1}\right)_{d_1,d_2} = p_0 p_1 \int_0^1 \frac{d H_{d_1}(v)}{dv}\Bigg|_{v=u} \frac{d H_{d_2}(v)}{dv}\Bigg|_{v=u}  du \, ,
\end{equation*}
with $H_d(u) = f_{1}(Q_{1}(v)) \partial_{\theta_{d}} G(Q_{0}(v),\theta_0)$. Next, proceeding similarly as in ACM, we conclude that:

\begin{equation*}
    \left(\mathbbm{V}^{*-1}\right)_{d_1,d_2} = p_0 p_1 (I(\theta_0))_{d_1,d_2} \, ,
\end{equation*}
where $I(\theta_0)$ is the Fisher information matrix of the parametric model $\theta \mapsto f_{1}(y|\theta)$ that assumes $Q_{0}$ known. It then follows by Lemma 1 of \cite{Athey2023} that the estimator is asymptotically efficient, as it achieves the efficiency bound derived by the authors.

To conclude, we note that, since the $\{P_l\}_l$ form an orthonormal \textbf{basis}, the estimator $\check{\theta}$ corresponds to GMLM that uses \textbf{infinitely} many moments. We are then able to show, by reasoning similarly as in Appendix G of ACM, that an estimator \eqref{eq_estimator} that uses a finite but increasing number of L-moments and a suitable choice of weights, is also efficient, since under an appropriate choice of weights this estimator is asymptotically equivalent to $\check{\theta}$. But then, since the optimally weighted GMLM estimator minimises, for each $R$, the variance of the leading term of the approximation \eqref{asy_strong}, we conclude that it is also efficient.

\label{app_efficiency}
\subsection{Selecting the number of L-moments} In a parametric setting, ACM propose two methods to select the number of L-moments. One method relies on higher-order expansions of the estimator \citep{Donald2009}; whilst the other on regularised estimation of the moment-combination matrix \citep{Luo2016}. Similarly, \cite{Athey2023} propose a method to choose the trimming constants $(\underline{p}, \overline{p})$ based on minimization of an estimate of the asymptotic variance of their estimator.\footnote{See \cite{Crump2009} on an earlier example of this approach in the context of observational studies.}

While these different approaches can be extended to our semiparametric estimator, we note that, in an experimental setting, both rely on tuning hyperparameters in the same sample used in the analysis. Inasmuch as, in finite samples, there is some room for overfitting, such approaches may somewhat hinder the credibility of the analysis, going counter to one of the primary goals of employing randomization in the first place \citep{athey2017econometrics}. 

Motivated by these concerns, and inspired by the literature on synthetic controls \citep{abadie2021using}, we propose a method to tune $(R,\underline{p},\overline{p})$ which relies on availability of pre-treatment data for the units in the experimental sample. Specifically, we assume that, for each unit in the sample, we observe the outcome of interest $Y$ for pre-treatment periods $t \in \{-1,-2,\ldots, -T_0\}$. In addition, we assume that the semiparametric model \eqref{eq_model} admits a ``null effect'' parameter value, i.e. there exists a known $\tilde{\theta} \in \Theta$ such that:

$$G(y;\tilde{\theta}) = y, \quad \forall y \in \mathbb{R} \, .$$

For each $t \in \{-1,-2,\ldots, - T_{0}\}$, let $\hat{\theta}(t;h)$ be the ``placebo'' GMLM estimator that replaces $\hat{Q}_{d}$ with the empirical quantiles of the outcome at the pre-treatment period $t$ in the group whose post-treatment status is $d$. We denote by $h$ the set of hyperparameters used in the estimation. Under the assumption that the treatment is not anticipated, we would expect $\hat{\theta}(t;h)$ to be close to $\tilde{\theta}$. Consequently, letting $\mathbf{H}$ denote a grid for the hyperparameters, we propose to choose the hyperparameters $h^*$ by minimizing:

$$h^* \operatorname{argmin}_{h \in \mathbf{H}} \frac{1}{T_0} \sum_{t = -1}^{-T_0} \lVert \hat{\theta}(t;h) - \tilde{\theta}\rVert_2^2 \, .$$
\label{app_selection} 

Our approach to calibrating the hyperparameters amounts to choosing those that entail the best pre-treatment fit, where ``fit'' is to be understood as minimising the average squared Euclidean distance between the estimated model parameters and the ``null effect'' parameter value. Notice that, with a large of number of pre-treatment periods, and assuming a law of large numbers holds at the time series, our approach to calibration may be seen as minimising average mean-squared error. 

\subsection{Monte Carlo exercise} \label{app_montecarlo} We revisit an exercise in \cite{Athey2023}, which is based on house price data available in \cite{Linden2008}. Following \cite{Athey2023}, we independently draw two samples of size $\frac{N}{2}$ without replacement from the price data, and label one of these as the treatment group. We contrast the behaviour of four different estimators in estimating the (zero) population average treatment effect: the difference-in-means estimator, \citeauthor{Athey2023}'s efficient influence function (EIF) estimator, and the optimally-weighted semiparametric generalised L-moment estimator, with and without (upper-tail) trimming. To select the hyperparameters in the GMLM, we generate $T_0 = 16$ pre-treatment periods by independently drawing from the price data.\footnote{The value of $T_0$ is chosen to more closely replicate the empirical application in the main text, where $19$ pre-treatment periods are available.} We consider $N \in \{500, 1000, 2000\}$, and, for each sample size, run $1,000$ replications.

Table \ref{mc_level} reports results from employing the semiparametric location model $Y(1) = \theta_0 + Y(0)$. As we can see, the semiparametric methods are able to provide substantial improvements over the nonparametric approach, both in terms of root mean squared error and mean absolute error, as well as shorter confidence intervals. The GMLM with no trimming offers further improvement in estimation error over the EIF approach, with reductions in the root mean squared error ranging from 50 ($N=2000$) to 180 ($N=500$) dollars. Confidence intervals have coverage close to the nominal level. In this setting, trimming does not appear to offer significant improvements: on the contrary, when employing trimming, the J-test appears to slightly overreject, whereas the non-trimmed version is closer to the nominal level.

Table \ref{mc_log} reports results from working with the data in logs. In this case, we employ the semiparametric model $\log(Y(1)) = \theta_0 + \log(Y(0))$. In this setting, both the difference-in-means and the semiparametric methods have a similar performance. The J-test has rejection rate close to the nominal level for both the trimmed and non-trimmed GMLM.

    % latex table generated in R 4.3.1 by xtable 1.8-4 package
% Mon Oct 23 15:37:21 2023
\begin{table}[H]
\centering
\caption{Monte Carlo: model in levels} 
\label{mc_level}
\scalebox{0.8}{
\begin{tabular}{lllllll}
   \hline \\ & & & \multicolumn{2}{c}{95\% CI} & &\\ \cmidrule(lr){4-5} & RMSE & MAE & Coverage & Average length & J-test & Median R \\ 
   \hline
\textit{N = 500 } &  &  &  &  &  &  \\ 
  Difference in means & 11910.56 & 9614.08  & 0.95     & 46299.88 &          &          \\ 
  Efficient Influence Function & 6096.84  & 4893.39  & 0.98     & 27873.95 &          &          \\ 
  L-moment estimator (no trimming) & 5916.71  & 4720.83  & 0.94     & 22041.1  & 0.06     & 7        \\ 
  L-moment estimator (trimming) & 5928.17  & 4738.4   & 0.94     & 22095.1  & 0.08     & 7        \\ 
   \hline
\textit{N = 1000 } &  &  &  &  &  &  \\ 
  Difference in means & 8340.56  & 6623.62  & 0.95     & 32743.11 &          &          \\ 
  Efficient Influence Function & 4324.92  & 3485.73  & 0.97     & 18901.91 &          &          \\ 
  L-moment estimator (no trimming) & 4253.06  & 3436.87  & 0.93     & 15112.15 & 0.06     & 8        \\ 
  L-moment estimator (trimming) & 4244.89  & 3430.02  & 0.92     & 15134.11 & 0.09     & 8        \\ 
   \hline
\textit{N = 2000 } &  &  &  &  &  &  \\ 
  Difference in means & 5944.96  & 4756.24  & 0.95     & 23163.82 &          &          \\ 
  Efficient Influence Function & 3003.63  & 2398.17  & 0.97     & 12925.77 &          &          \\ 
  L-moment estimator (no trimming) & 2951.57  & 2339.18  & 0.94     & 11267.76 & 0.03     & 8        \\ 
  L-moment estimator (trimming) & 2937.56  & 2328.19  & 0.94     & 11187.06 & 0.08     & 8        \\ 
   \hline
\end{tabular}
}

 \begin{minipage}{0.8\linewidth}
 \vspace{1em}
%       \begin{minipage}{\textwidth} -- what should go in this width argument?
            \footnotesize \textit{Notes:} outcomes are measured in dollars. The number of replications is set to $1,000$. RMSE stands for root-mean squared error. MAE stands for mean absolute error. The confidence intervals are based on a normal approximation and on an estimate of the estimator's asymptotic variance. Column \textit{J-test} reports the rejection rate of a test of overidentifying-restrictions based on the J-statistic \eqref{eq_jtest} at the 5\% level. 
        \end{minipage}
\end{table}

    % latex table generated in R 4.3.1 by xtable 1.8-4 package
% Mon Oct 23 15:37:15 2023
\begin{table}[H]
\centering
\caption{Monte Carlo: model in logs} 
\label{mc_log}
\scalebox{0.8}{
\begin{tabular}{lllllll}
   \hline \\ & & & \multicolumn{2}{c}{95\% CI} & &\\ \cmidrule(lr){4-5} & RMSE & MAE & Coverage & Average length & J-test & Median R \\ 
   \hline
\textit{N = 500 } &  &  &  &  &  &  \\ 
  Difference in means & 0.05 & 0.04 & 0.95 & 0.21 &      &      \\ 
  Efficient Influence Function & 0.04 & 0.03 & 0.96 & 0.17 &      &      \\ 
  L-moment estimator (no trimming) & 0.04 & 0.04 & 0.94 & 0.16 & 0.05 & 6    \\ 
  L-moment estimator (trimming) & 0.04 & 0.04 & 0.94 & 0.16 & 0.05 & 6    \\ 
   \hline
\textit{N = 1000 } &  &  &  &  &  &  \\ 
  Difference in means & 0.04 & 0.03 & 0.94 & 0.15 &      &      \\ 
  Efficient Influence Function & 0.03 & 0.02 & 0.96 & 0.12 &      &      \\ 
  L-moment estimator (no trimming) & 0.03 & 0.02 & 0.94 & 0.11 & 0.05 & 6    \\ 
  L-moment estimator (trimming) & 0.03 & 0.02 & 0.93 & 0.11 & 0.06 & 6    \\ 
   \hline
\textit{N = 2000 } &  &  &  &  &  &  \\ 
  Difference in means & 0.03 & 0.02 & 0.95 & 0.1  &      &      \\ 
  Efficient Influence Function & 0.02 & 0.02 & 0.95 & 0.08 &      &      \\ 
  L-moment estimator (no trimming) & 0.02 & 0.02 & 0.95 & 0.08 & 0.04 & 6    \\ 
  L-moment estimator (trimming) & 0.02 & 0.02 & 0.94 & 0.08 & 0.05 & 7    \\ 
   \hline
\end{tabular}
}

 \begin{minipage}{0.8\linewidth}
 \vspace{1em}
%       \begin{minipage}{\textwidth} -- what should go in this width argument?
            \footnotesize \textit{Notes:} outcomes are measured in logs. The number of replications is set to $1,000$. RMSE stands for root-mean squared error. MAE stands for mean absolute error. The confidence intervals are based on a normal approximation and on an estimate of the estimator's asymptotic variance. Column \textit{J-test} reports the rejection rate of a test of overidentifying-restrictions based on the J-statistic \eqref{eq_jtest} at the 5\% level. 
        \end{minipage}
\end{table}

\section{Additional figures}
\label{app_figures}
\begin{figure}
    \centering
     \begin{subfigure}{\textwidth}
     \centering
     \caption{Average}
     \includegraphics[scale=0.45]{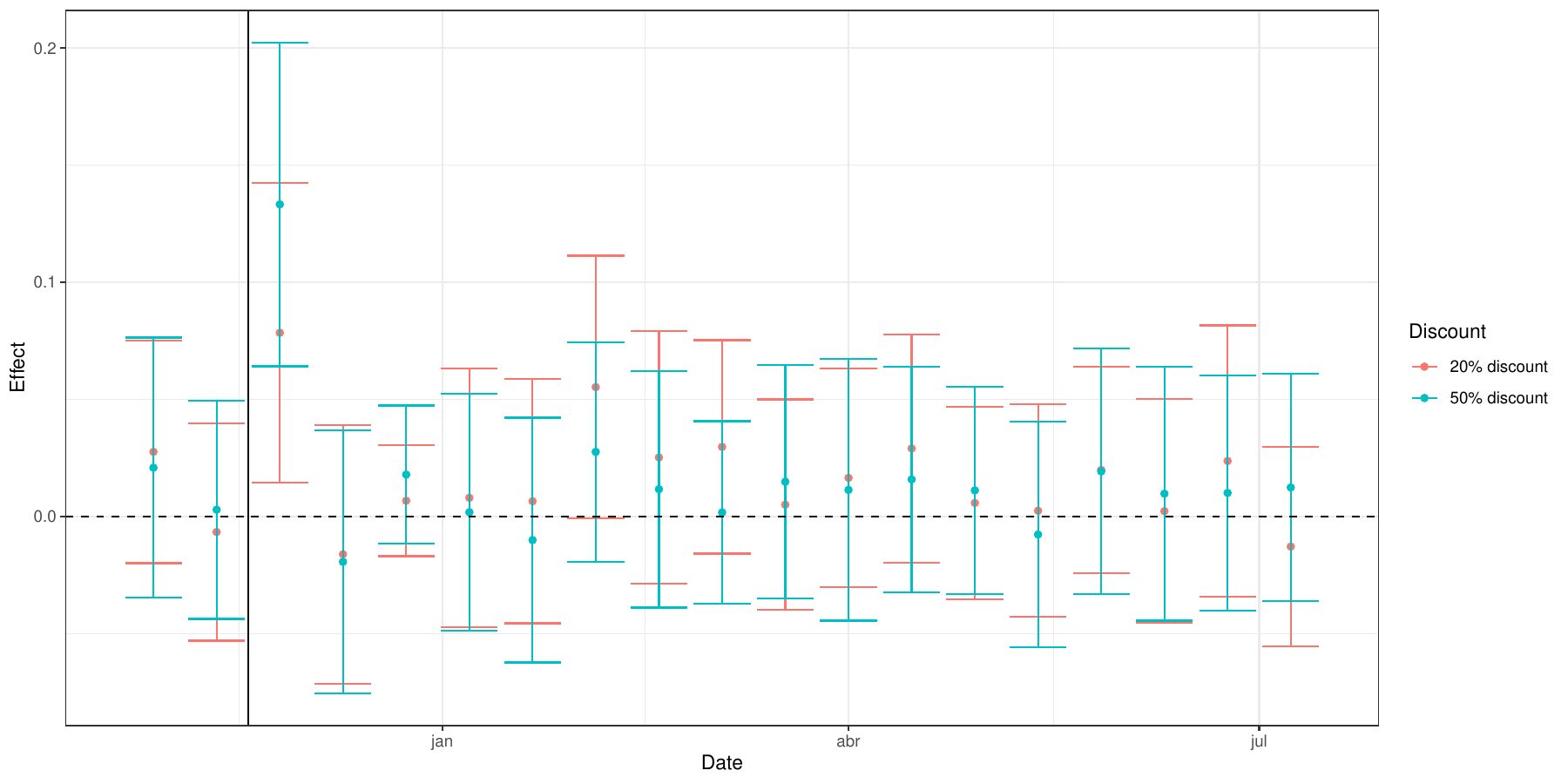}
     \end{subfigure}
          \begin{subfigure}{\textwidth}
     \centering
     \caption{Dispersion}
     \includegraphics[scale=0.45]{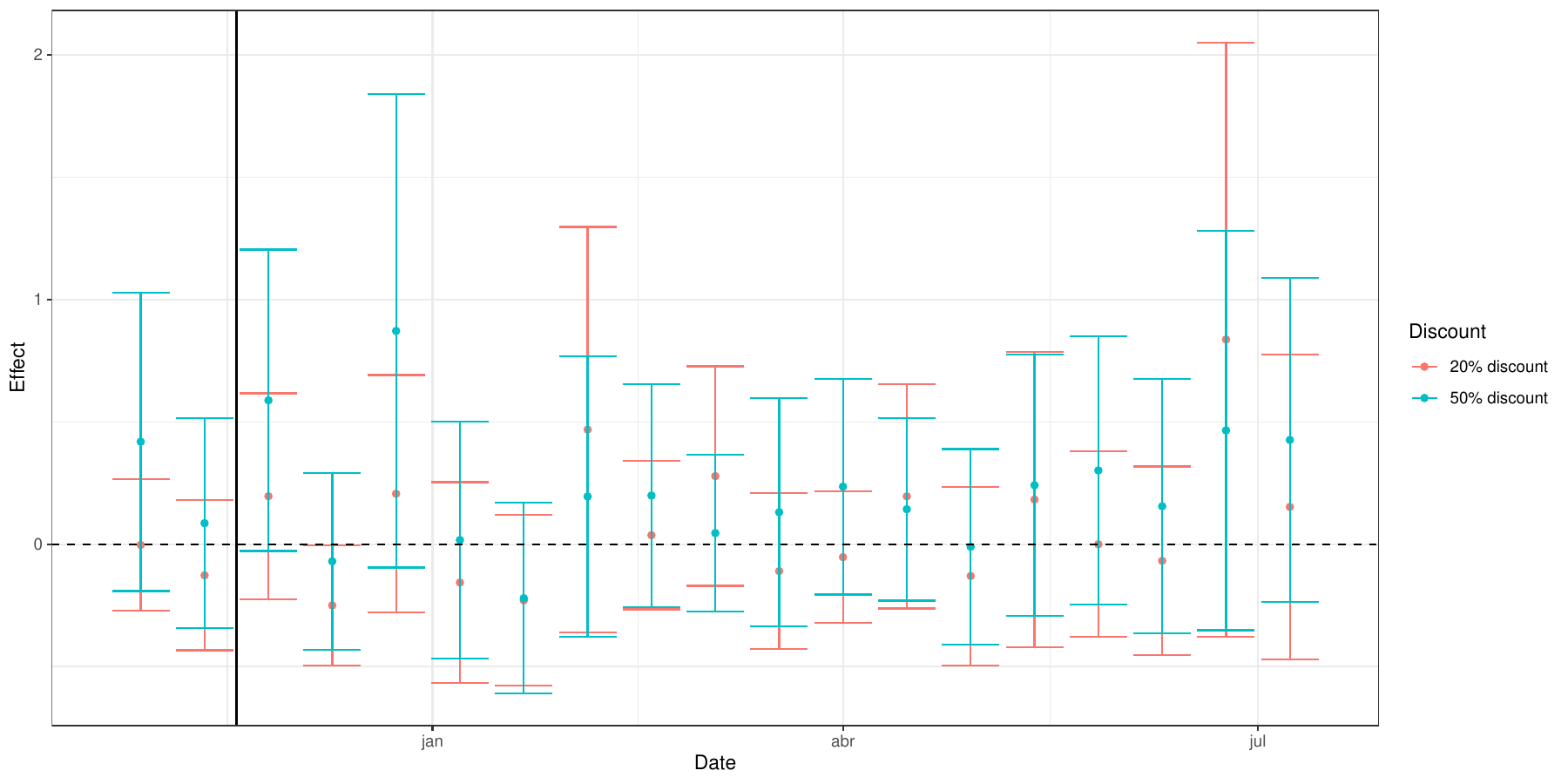}
     \end{subfigure}
    \caption{Nonparametric estimates of effects of discounts on integrated rides}
    \label{fig:bimodal_np}

     \begin{minipage}{0.8\linewidth}
 \vspace{1em}
%       \begin{minipage}{\textwidth} -- what should go in this width argument?
            \footnotesize \textit{Notes:}
        The figure plots estimates, along with 95\% confidence intervals, of the effects of each discount on the mean and dipersion of integrated rides across the last two pre-treament fortnights (left of vertical dark line) and the post-treatment window. We aggregate effects across prior train/metro usage by averaging the nonparametric estimators discussed in the main text across $u$, with weights given by the proportion of usage type $u$ in the sample. Confidence intervals are computed using a normal approximation and the delta method. For reference, the average number of integrated rides in the control (no-discount) group in each fortnight is: \textit{2018-10-29}: 0.12; \textit{2018-11-12}: 0.11; \textit{2018-11-26}: 0.13; \textit{2018-12-10}: 0.14; \textit{2018-12-24}: 0.04; \textit{2019-01-07}: 0.11; \textit{2019-01-21}: 0.11; \textit{2019-02-04}: 0.08; \textit{2019-02-18}: 0.12; \textit{2019-03-04}: 0.1; \textit{2019-03-18}: 0.1; \textit{2019-04-01}: 0.12; \textit{2019-04-15}: 0.11; \textit{2019-04-29}: 0.08; \textit{2019-05-13}: 0.1; \textit{2019-05-27}: 0.09; \textit{2019-06-10}: 0.11; \textit{2019-06-24}: 0.11; \textit{2019-07-08}: 0.1.
    \end{minipage}
    
\end{figure}

\begin{figure}[H]
    \centering
     \begin{subfigure}{\textwidth}
     \centering
     \caption{Average}
     \includegraphics[scale=0.45]{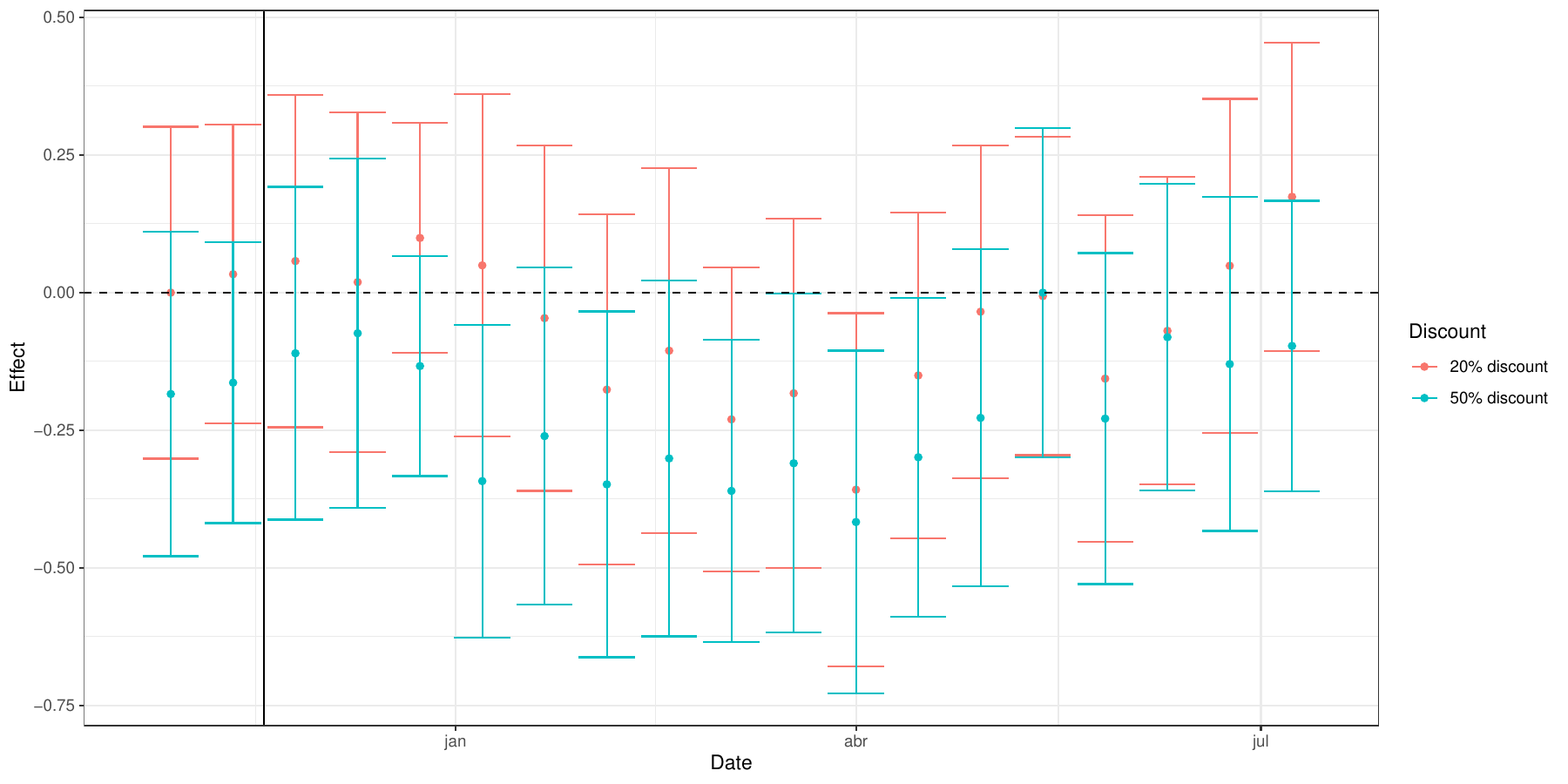}
     \end{subfigure}
          \begin{subfigure}{\textwidth}
     \centering
     \caption{Dispersion}
     \includegraphics[scale=0.45]{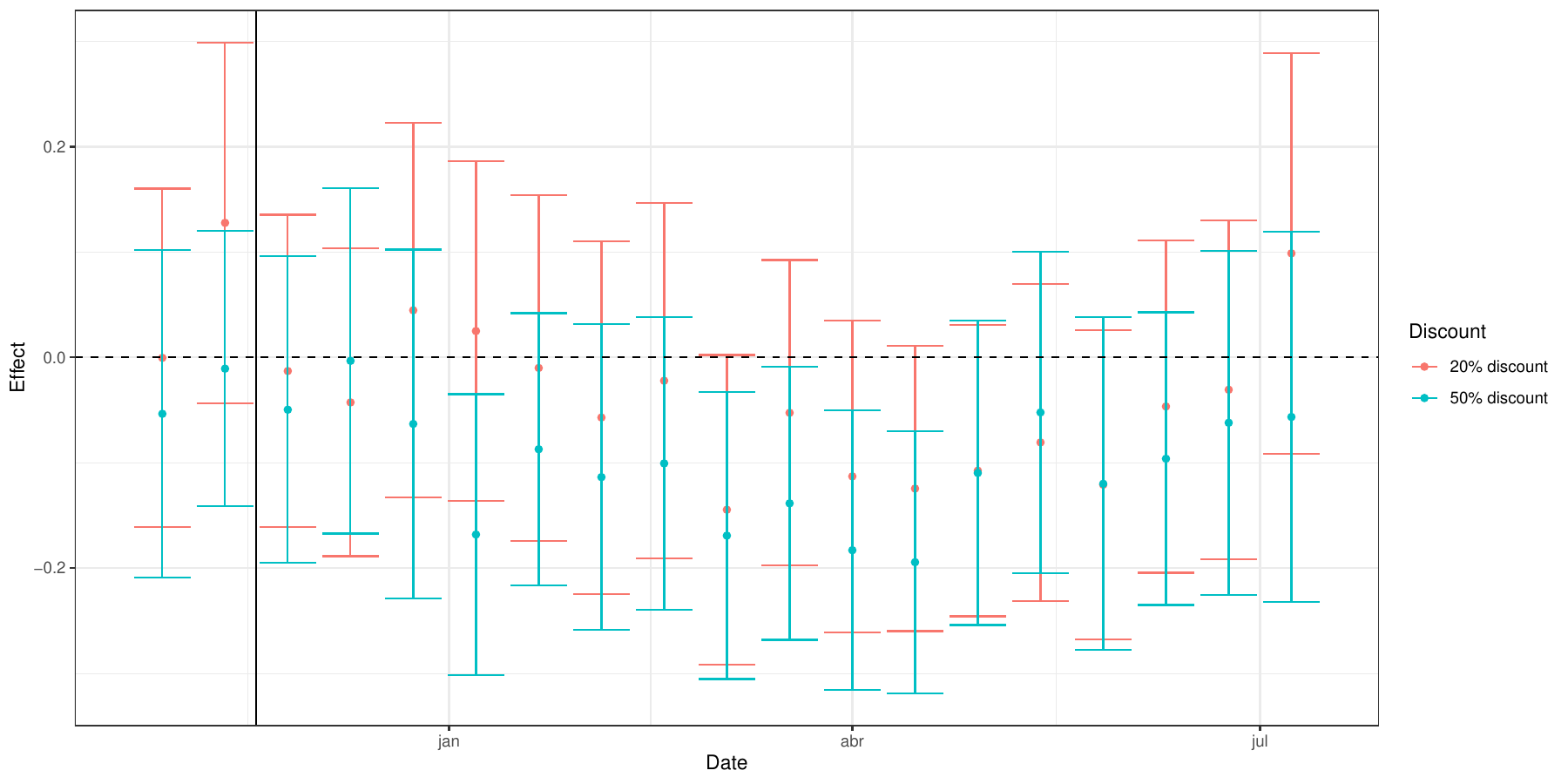}
     \end{subfigure}
    \caption{Nonparametrics estimates of discounts on non-integrated rides}
    \label{fig:unimodal_np}

 \begin{minipage}{0.8\linewidth}
 \vspace{1em}
%       \begin{minipage}{\textwidth} -- what should go in this width argument?
            \footnotesize \textit{Notes:}
        The figure plots estimates, along with 95\% confidence intervals, of the effects of each discount on the mean and dipersion of non-integrated rides across the last two pre-treament fortnights (left of vertical dark line) and the post-treatment window. We aggregate effects across prior train/metro usage by averaging the nonparametric estimators discussed in the main text across $u$, with weights given by the proportion of usage type $u$ in the sample. Confidence intervals are computed using a normal approximation and the delta method. For reference, the average number of non-integrated rides in the control (no-discount) group in each fortnight is: \textit{2018-10-29}: 2.14; \textit{2018-11-12}: 2.02; \textit{2018-11-26}: 2.23; \textit{2018-12-10}: 2.15; \textit{2018-12-24}: 1.24; \textit{2019-01-07}: 1.96; \textit{2019-01-21}: 2.11; \textit{2019-02-04}: 2.11; \textit{2019-02-18}: 2.27; \textit{2019-03-04}: 1.9; \textit{2019-03-18}: 2.05; \textit{2019-04-01}: 2.14; \textit{2019-04-15}: 1.97; \textit{2019-04-29}: 1.88; \textit{2019-05-13}: 1.73; \textit{2019-05-27}: 1.84; \textit{2019-06-10}: 1.74; \textit{2019-06-24}: 1.81; \textit{2019-07-08}: 1.52.
    \end{minipage}
\end{figure}

\begin{figure}
    \centering
     \begin{subfigure}{0.4\textwidth}
     \centering
     \caption{Bimodal}
     \includegraphics[scale=0.45]{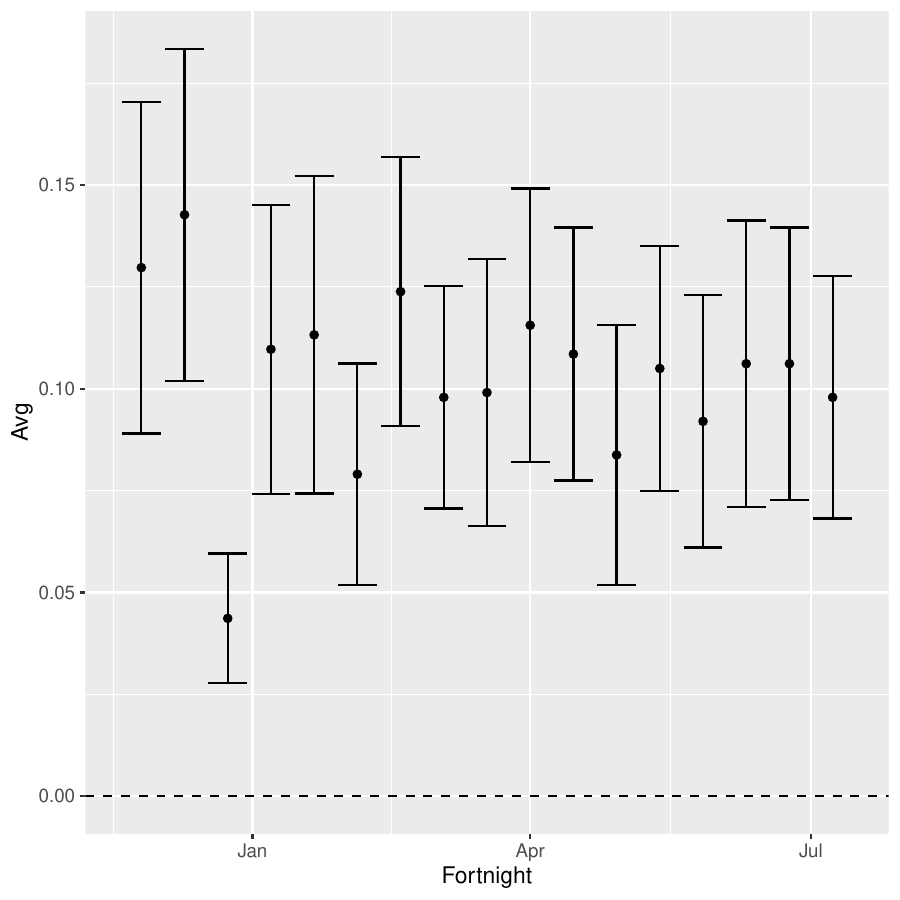}
     \end{subfigure} \begin{subfigure}{0.4\textwidth}
     \centering
     \caption{Unimodal}
     \includegraphics[scale=0.45]{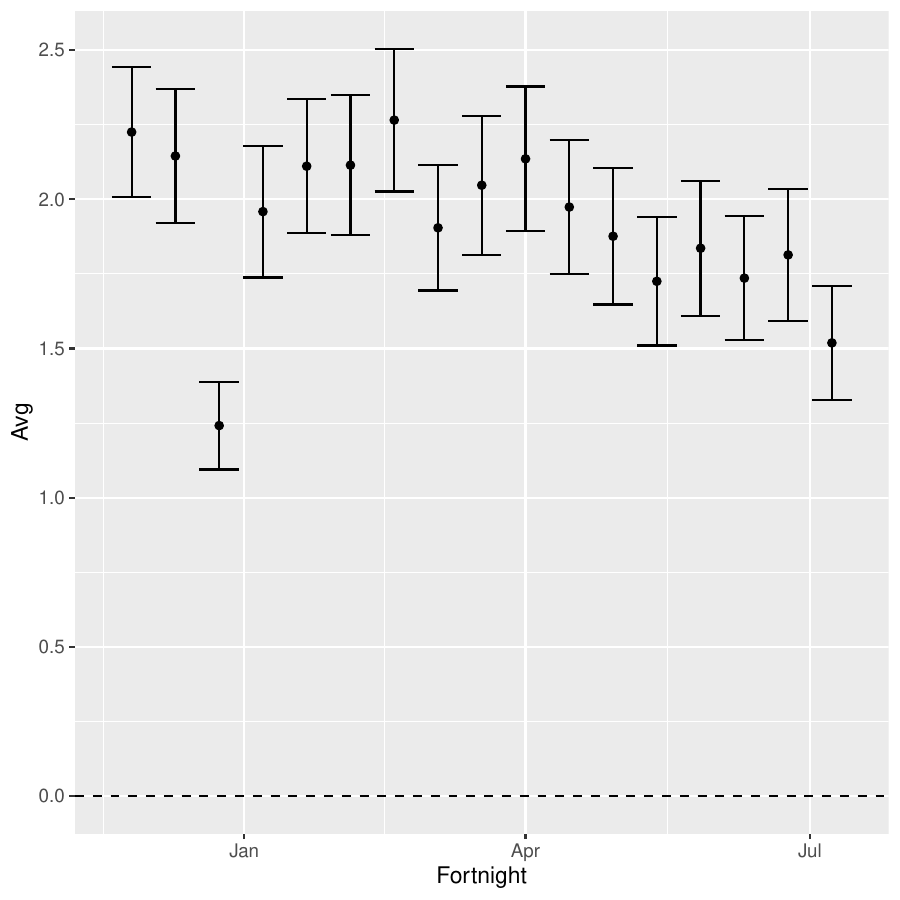}
     \end{subfigure}
    \caption{Average number of rides in the control group in the post-treatment window, along with 95\% confidence intervals}
    \label{fig:avg_control}
\end{figure}

Figures \ref{fig:bimodal_np} and \ref{fig:unimodal_np} replicate Figures \ref{fig:bimodal_text} and \ref{fig:unimodal_text} in the main text, but replacing the semiparametric estimators used in each aggregation with the corresponding nonparametric estimators reported in Tables \ref{table_not_user_bimodal}, \ref{table_user_bimodal}, \ref{table_not_user_unimodal}, and \ref{table_user_unimodal}. Figure \ref{fig:avg_control} reports the average number of each type of ride in the control group, along with 95\% confidence intervals, in the posstreatment window.

\begin{landscape}
	\section{Baseline survey data}
	Table \ref{table_data_raw} reports the variables collected at the baseline survey. In addition to this information, and as discussed in the main text, we observe, for each individual who answered the survey, the number of integrated and nonintegrated rides taken between early 2018 until mid-2019, on a biweekly basis. The latter piece of information was collected by the ridesharing company's staff from each user's history in the ridesharing platform.
	
\begin{table}[H]
	\caption{Details on data collected at the baseline survey}
	\label{table_data_raw}
	\scalebox{0.54}{
		\begin{tabular}{c|c|c}
			\textbf{Variable Label} & \textbf{Description} & \textbf{Type} \\
			\hline hash\_userid & User id (99 attribute - hashed) & String \\
			\hline age & User age (declared) & Numeric (years) \\
			\hline gender & User gender (declared) & Factor: Feminino $=$ Weman; Masculino $=$ Men; Outro $=$ other \\
			\hline educ & User educational level (declared) & Factor \\
			\hline income & User income level (declared) & Factor: breaks in BRL \\
			\hline d\_dependents & Dummy: has dependents (declared) & Dummy: 1 = has dependents; 0 = has no dependents \\
			\hline dependents & Number of dependents (declared) & Numeric (number of dependents) \\
			\hline st home & Closest station to user's residence (declared) & String \\
			\hline st\_home\_clean & Closest station to user's residence (standardized) & String \\
			\hline d\_st home & Dummy: user consider distance between closest station and residence walkable (declareo & Dummy: 1 = walkable distance; $0=$ non-walkable distance \\
			\hline actixity & User economic activity (declared) & Factor \\
			\hline st work & Closest station to user's work (declared) & String \\
			\hline st\_ work\_clean & Closest station to user's work (standardized) & String \\
			\hline d\_st\_work & Dummy: user consider distance between closest station and work walkable (declared) & Dummy: 1 = walkable distance; $0=$ non-walkable distance \\
			\hline st\_study & Closest station to user's study place (declared) & String \\
			\hline st\_study\_clean & Closest station to user's study place (standardized) & String \\
			\hline d\_st\_study & Dummy: user consider distance between closest station and study place walkable (declare & Dummy: $1=$ walkable distance; $0=$ non-walkable distance \\
			\hline mode1\_work & Main mode of transpertation used to go/come to/from work (declared) & Factor \\
			\hline mode2\_work & Secondary mede of transpertation used to go/come to/from work (declared) & Factor \\
			\hline duration\_work & Time spent to go to work (declared) & Numeric (minutes) \\
			\hline mode1\_study & Main mode of transpertation used to go/come to/from study place (declared) & Factor \\
			\hline mode2\_study & Secondary mode of transpertation used to go/come to/from study place (declared) & Factor \\
			\hline duration\_study & Time spent to go to study place (declared) & Numeric (minutes) \\
			\hline disbursement\_workstudy & Monetary amount spent on transpertation (declared) & Factor \\
			\hline mode1\_others & Main mode of transportation ysed on other trips (declared) & Factor \\
			\hline disbursement others & Monetary amount spent on other trips (declared) & Factor \\
			\hline start\_date & Survey start date/time & Date/time \\
			\hline submit\_date & Suryey finish date/time & Date/time \\
			\hline network id & Survey network id & String \\
			\hline
		\end{tabular}
	}
\end{table}
\end{landscape}

 \end{document}